\documentclass[nofootinbib,prx,twocolumn,superscriptaddress,citeautoscript,amsart]{revtex4-2}

\usepackage{graphicx}
\usepackage{multirow}
\usepackage{color}
\usepackage{bm}
\usepackage{times}
\usepackage{amsmath,bm,amsfonts}
\usepackage{dcolumn}
\usepackage{graphicx}
\usepackage{latexsym}
\usepackage{hhline}
\usepackage{braket}
\usepackage{bbold}

\def\ket#1{|#1\rangle }

\def\bb{\mathbb}

\def\d{\partial}

\begin{document}
\title{
Low-frequency divergence and quantum geometry of the bulk photovoltaic effect in topological semimetals}
\author{Junyeong \surname{Ahn}}
\email{Present address: 
Department of Physics, Harvard University, Cambridge, Massachusetts 02138, USA,
junyeongahn@fas.harvard.edu}
\affiliation{RIKEN Center for Emergent Matter Science (CEMS), Wako, Saitama 351-0198, Japan}
\affiliation{Department of Applied Physics, The University of Tokyo, Bunkyo, Tokyo 113-8656, Japan}

\author{Guang-Yu \surname{Guo}}
\email{gyguo@phys.ntu.edu.tw}
\affiliation{Department of Physics and Center for Theoretical Physics, National Taiwan University, Taipei 10617, Taiwan}
\affiliation{Physics Division, National Center for Theoretical Sciences, Hsinchu 30013, Taiwan}

\author{Naoto \surname{Nagaosa}}
\email{nagaosa@riken.jp}
\affiliation{RIKEN Center for Emergent Matter Science (CEMS), Wako, Saitama 351-0198, Japan}
\affiliation{Department of Applied Physics, The University of Tokyo, Bunkyo, Tokyo 113-8656, Japan}

\date{\today}

\begin{abstract}
We study the low-frequency properties of the bulk photovoltaic effect in topological semimetals.
The bulk photovoltaic effect is a nonlinear optical effect that generates DC photocurrents under uniform irradiation, allowed by noncentrosymmetry.
It is a promising mechanism for a terahertz photodetection based on topological semimetals.
Here, we systematically investigate the low-frequency behavior of the second-order optical conductivity in point-node semimetals.	
Through symmetry and power-counting analysis, we show that Dirac and Weyl points with tilted cones show the leading low-frequency divergence.
In particular, we find new divergent behaviors of the conductivity of Dirac and Weyl points under circularly polarized light, where the conductivity scales as $\omega^{-2}$ and $\omega^{-1}$ near the gap-closing point in two and three dimensions, respectively.
We provide a further perspective on the low-frequency bulk photovoltaic effect by revealing the complete quantum geometric meaning of the second-order optical conductivity tensor.
The bulk photovoltaic effect has two origins, which are the transition of electron position and the transition of electron velocity during the optical excitation, and the resulting photocurrents are respectively called the shift current and the injection current.
Based on an analysis of two-band models, we show that the injection current is controlled by the quantum metric and Berry curvature, whereas the shift current is governed by the Christoffel symbols near the gap-closing points in semimetals.
Finally, for further demonstrations of our theory beyond simple two-band models,
we perform first-principles calculations on the shift and injection photocurrent conductivities as well as geometric
quantities of antiferromagnetic MnGeO$_3$ and ferromagnetic PrGeAl, respectively, as representatives of real magnetic Dirac and Weyl semimetals.
Our calculations reveal gigantic peaks in many nonvanishing elements of photoconductivity tensors below photon energy $\sim$0.2 eV 
in both MnGeO$_3$ and PrGeAl.
In particular, we show the $\omega^{-1}$ enhancement of the shift conductivity tensors due to the divergent behavior of the geometric quantities 
near the Dirac and Weyl points as well as slightly gapped topological nodes.  
Moreover, the low-frequency bulk photovoltaic effect is tunable by carrier doping and magnetization orientation rotation.
Our work brings out new insights into the structure of nonlinear optical responses as well as for the design of semimetal-based terahertz photodetectors.
\end{abstract}

\maketitle

\section{Introduction}

Topological semimetals are emerging as efficient infrared and terahertz photodetectors~\cite{liu2020semimetals}.
In contrast to semiconductors whose absorption spectum is bounded below by the band gap, semimetals can detect radiations down to the terahertz range because of their gapless spectrum.
A promising mechanism for the generation of photocurrents in semimetals is the bulk photovoltaic effect.
It refers to the generation of photocurrent under uniform irradiation of light due to the intrinsic inversion asymmetry of the system.
Since the bulk photovoltaic effect does not require a bias voltage for breaking inversion symmetry, dark current noise can be suppressed~\cite{liu2020semimetals}.

To achieve high photosensitivity, we need to understand how to obtain large photoconductivity.
It is believed that band topology plays an important role~\cite{kraut1979anomalous,von1981theory,morimoto2016topological,de2017quantized,nagaosa2017concept}.
The bulk photovoltaic effect occurs due to the inversion-asymmetric transition of electron position or velocity during the optical excitation, and the resulting photocurrents are respectively called the shift current and the injection current~\cite{sipe2000second}.
In nonmagnetic systems, linearly polarized light induces shift currents while circularly polarized light induces injection currents.
Remarkably, both the linear shift~\cite{morimoto2016topological} and circular injection~\cite{de2017quantized} currents were found to be intimately related to the topological quantities, the Berry connection and the Berry curvature, respectively.
These discoveries have led to various theoretical and experimetnal studies searching for topological enhancement near the gap-closing points~\cite{morimoto2016semiclassical,konig2017photogalvanic,kim2017shift,ma2017direct,wu2017giant,yang2017divergent,golub2018circular,flicker2018chiral,zhang2018photogalvanic,osterhoudt2019colossal,ma2019nonlinear,parker2019diagrammatic}.

However, while there is a concrete proportional quantitative relationship between the injection current and the Berry curvature, no such a quantitative relation exists between the shift current and the Berry connection.
For example, a Dirac point in two dimensions has a quantized $\pi$ Berry phase, thus having a nontrivial Berry connection.
Nevertheless, such a Dirac point does not generate a shift current because of its inversion symmetry.
Furthermore, no simple quantitative relation was found between the shift current and the shift vector~\cite{young2012first}, a gauge-invariant quantity related to the Berry connection, without some special requirements like the momentum independence of the dipole matrix elements~\cite{fregoso2017quantitative}.

The bulk photovoltaic effect in magnetic topological semimetals is more poorly understood, although a recent work has revealed some general aspects~\cite{holder2020consequences}.
Due to time-reversal symmetry breaking in magnetic systems, linearly (circularly) polarized light can generate injection (shift) currents as well as shift (injection) currents.
There are some works highlighting the generation of large linear injection currents in magnetic systems~\cite{semenov2012tunable,ogawa2016zero,chan2017photocurrents,zhang2019switchable,sun2019giant,holder2020consequences}, but the relationship between the response and band topology has not been understood.
Moreover, there has been very little attention to the circular shift current, while the first concept of the shift current appeared as a response to circularly polarized light~\cite{mendis1968photocurrent,genkin1968contribution,sturman1992photovoltaic}.
One reason for this is that injection currents are typically stronger than the shift currents.
Since there is a rapid progress in the experimental observation~\cite{liu2019magnetic,belopolski2019discovery,morali2019fermi} and theoretical proposal~\cite{xu2011chern,hirschberger2016chiral,kushwaha2018magnetic,zou2019study,jin2019noncentrosymmetric,tang2016dirac,armitage2018weyl,lin2019dirac,xu2020high} of various magnetic topological semimetals, addressing optical properties in magnetic topological semimetals is now a timely subject.

\begin{table*}[ht]
\begin{tabular}{c|cc|cc}
Response			& Linear Injection	& Circular Injection	& Linear Shift	& Circular Shift \\
\hhline{=|==|==}
Parity under $T$
&$-$
&$+$
&$+$
&$-$\\
Parity under $PT$
&$+$
&$-$
&$-$
&$+$\\
Geometric quantities
&Quantum metric
&Berry curvature
&Symplectic Christoffel symbols
&Christoffel symbols of the first kind\\
Leading divergence
&\multicolumn{2}{c|}{$O(\tau \omega^{d-3})$}
&\multicolumn{2}{c}{$O(\omega^{d-4})$}
\end{tabular}
\caption{
Low-frequency properties of second-order DC photovoltaic responses.
Linear and circular indicates the polarization of light.
Injection and shift conductivities are defined by Eq.~\eqref{eq:shift-inj}.
The sign $\pm$ indicates the parity of the conductivity tensor under the action of time reversal $T$ and spacetime inversion $PT$.
See Eqs.~\eqref{eq:linear-transform} and~\eqref{eq:circular-transform}.
In $T$- ($PT$-) symmetric systems, only the responses with the positive $T$- ($PT$-) parity appears.
All of the four responses can appear when $T$ and $PT$ symmetries are both broken.
Based on their symmetry properties, we call the linear injection and circular shift current responses as $PT$-symmetric responses and the linear shift and circular injection current responses as $T$-symmetric responses.
Symplectic Christoffel symbols indicate the symplectic analog of the Christoffel symbol of the first kind (See Eq.~\eqref{eq:symplectic-symbol}).
Here we consider only interband-transitive processes in the clean limit where the relaxation rate $\Gamma=\tau^{-1}$ is smaller than the photon frequency $\omega$.
In this case, the injection current is typically larger than the shift current, and they become comparable when $\omega$ approaches $\Gamma$.
}
\label{tab:symmetry}
\end{table*}

In this work, we reveal general low-frequency properties of the shift and injection currents in magnetic and nonmagnetic point-node semimetals.
By employing symmetry and power-counting analysis, we determine the leading low-frequency divergence near the gap-closing point, as summarized in Table~\ref{tab:symmetry}.
We show that tilted Dirac and Weyl points can generate the leading divergence.
While the bulk photovoltaic response in tilted Dirac and Weyl points have been studied~\cite{semenov2012tunable,ogawa2016zero,chan2017photocurrents,yang2017divergent}, here we approach them in a unified view and discover new aspects.
Our theory covers type-I and type-II spectrum of Dirac and Weyl points in any dimensions.
In particular, our analysis include type-II Dirac points in two dimensions and type-I and type-II Dirac points in three dimensions, which are relevant to magnetic Dirac semimetals, whose shift and injection currents have not been previously studied.
It is widely known that the protection of Dirac points against opening the gap requires inversion symmetry, which forbids the bulk photovoltaic response.
That is true in nonmagnetic systems.
However, in magnetic systems, symmetry under the combination of spatial inversion $P$  and time reversal $T$, which is $PT$ symmetry, is enough for the protection~\cite{armitage2018weyl,tang2016dirac,lin2019dirac,xu2020high}, so inversion symmetry can be broken.
Also, we study the largely overlooked circular shift current in Weyl and Dirac systems.
The circular shift current grows fast as the photon frequency gets smaller, and it scales as $\omega^{-1}$ in three dimensions.
This indicates that Dirac semimetals in three dimensions can show divergent photovoltaic responses like Weyl semimetals, although the Berry curvature is identically zero due to $PT$ symmetry.
In two dimensions, it grows faster as $\omega^{-2}$.

We find that tilted Dirac and Weyl points show what we call the {\it separation of responses}, where photocurrents of different origins manifest through different current directions.
For example, circular shift and circular injection currents flow in different directions.
This can be useful in the detection of shift currents in the coexistence of the stronger injection currents.
The separation of responses can occur from the symmetry transformation property under magnetic operations $MT$ or $C_{2}T$, where $M$ and $C_{2}$ are mirror and twofold rotation, so it remains robust in the system beyond the ${\bf k}$-linear approximation as long as those symmetries are preserved.

Our symmetry and power-counting analysis are enough for understanding the general pattern of the response for systems with a linear spectrum.
However, for a deeper understanding of the response, we propose a new perspective on the low-frequency bulk photovoltaic effect.
We uncover the full geometric nature of shift and injection currents.
Here, as well as the Berry curvature, another geometric quantity called the quantum metric has a crucial role.
Since geometry has more information than topology about quantum states, we observe the consequence of band topology in a more broad perspective through geometric quantities.
We show that the linear injection conductivity is determined only by the quantum metric near the gap closing, in the same way as the circular injection conductivity is determined by the Berry curvature.
This completes the geometric understanding of the injection currents at the low-frequency regime.
Furthermore, we show that the shift conductivities are related to a more interesting quantity, which is the {\it Christoffel symbols}.
Unlike the Berry connection, which has a complicated relationship to the magnitude of the shift current, the Christoffel symbols directly control the magnitude of the response.
In this viewpoint, the enhancement of the shift and injection current responses near the gap-closing point can be attributed to the divergence of the geometric quantities at the geometric singularity, i.e., the gap-closing point.
Furthermore, our unique perspective allows to view the bulk photovoltaic effect as novel tools for experimental measurements of quantum geometry in materials.

Our geometric interpretation of the bulk photovoltaic effect is clearly distinguished from the one in Ref.~\cite{holder2020consequences}, which also discuss the role of the quantum metric.
In Ref.~\cite{holder2020consequences}, the shift current is decomposed into four parts, and one of them was interpreted as the geometric contribution.
The injection current is identified to have no geometric origin.
These are in contrast to our interpretation in which both shift and injection currents are fully geometric responses.
Only through our direct quantitative relationships the bulk photovoltaic effect can be identified as a useful measure of quantum geometry in experiments as well as a way to theoretically understand the low-frequency divergent behavior.

Finally, for further demonstrations of our theory beyond simple two-band models,
we perform first-principles relativistic band theoretical calculations of the shift and injection photocurrent conductivities 
as well as geometric quantities in antiferromagnetic (AF) MnGeO$_3$ and ferromagnetic (FM) PrGeAl (as will be reported in Sec. VI below), 
respectively, as representatives of real magnetic Dirac and Weyl semimetals.
In AF MnGeO$_3$, although both $T$ and $P$ symmetries are broken, the combined $PT$ symmetry is preserved~\cite{xu2020high}.
Thus, AF MnGeO$_3$ was recently predicted to a rare magnetic Dirac semimetal.~\cite{xu2020high}
As our theory predicts (Table I), we find nonzero elements of circular shift and linear injection photoconductivity tensors in AF MnGeO$_3$.
In fact, several nonzero elements exhibit huge peaks below photon energy of $\sim$0.2 eV. 
Our calculations reveal that at least there are three Dirac points in the vicinity of the Fermi level
and two of the Dirac points are accompanied by a slightly gapped Dirac point each.
The calculated quantum metric and Christoffel symbol of first kind exhibit divergent behaviors near 
both the gapless and gapped Dirac points, thus leading to 
the geometric enhancement in
linear injection and circular shift currents, respectively.
In contrast, both $T$ and $PT$ symmetries are broken in FM PrGeAl~\cite{chang2018magnetic},
and hence all four types of the bulk photovoltaic effect may emerge in FM PrGeAl, as our theory predicts (Table I).
Indeed, we find that many nonzero elements of all shift and injection conductivity tensors
show gigantic peaks in the low-frequency range up to 0.1 eV. 
There are at least 160 type I and type II Weyl points within $\pm$0.1 eV of the Fermi level in FM PrGeAl~\cite{chang2018magnetic}.
Our calculations indicate that the low-frequency peaks can be further increased by up to a factor of 5
by raising the chemical potential from the Fermi level to the energy of certain Weyl points.
Our calculations also reveal the divergent behaviors of both quantum geometric tensor and quantum geometric connection
(symplectic Christoffel symbol and Christoffel symbol of first kind)
near the Weyl points and also anticrossing topological nodes, 
thus leading to the gigantic low-frequency shift and injection currents in FM PrGeAl.
Furthermore, we notice that FM PrGeAl is a soft ferromagnet~\cite{sanchez2020observation}.
Thus, the magnetization direction can be easily rotated away from the easy $c$-axis to,
e.g., $a$-axis, and this may cause a topology change of the Weyl point distribution 
in the Brilloin zone~\cite{chang2018magnetic}, thereby resulting in
significant changes in the shift and injection photoconductivities.
Also, since the circular shift and injection conductivity tensors 
are antisymmetric with respect to the magnetization direction,
they would change sign when the magnetization direction is reversed.
All these observations show that the gigantic low-frequency shift and injection photocurrents
in FM PrGeAl can be tuned by either carrier doping or magnetization direction rotation.

The outline of this paper is as follows.
We explain the shift and injection currents as a second-order optical response in Sec.~\ref{sec:shift-injection}.
Then, we study the symmetry and low-frequency divergence of the shift and injection currents in Sec.~\ref{sec:symmetry}.
Section~\ref{sec:geometry} enriches our analysis by revealing the quantum geometrical nature of the low-frequency responses.
We elaborate more on the symmetry and divergence with concrete models and numerical calculations in Sec.~\ref{sec:model}.
In Sec. VI, we present our first-principles calculations on the shift and injection photocurrent conductivities as well as geometric
quantities of AF MnGeO$_3$ and FM PrGeAl. In particular,
we analyze the gigantic peaks in the calculated low-frequency photoconductivity spectra 
in MnGeO$_3$ and PrGeAl in terms of the divergent behaviors of the geometric quantities
near the gapless and slightly gapped topological nodal points.
Finally, we discuss several issues about the low-frequency divergence 
in Sec.~\ref{sec:discussion}.

\section{Shift and Injection currents}
\label{sec:shift-injection}

Let us expand current density $j$ in increasing power of the electric field $E$ as
\begin{align}
\label{eq:expansion}
j^c
&=\sigma^{c;a}_{(1)}E_a+\sigma^{c;ab}_{(2)}E_aE_b+\sigma^{c;abc}_{(3)}E_aE_bE_c+\hdots.
\end{align}
The first term is the familiar linear response, and the other terms are nonlinear responses.
Since the current density oscillates in-phase with the electric field in the linear response, the DC photocurrent generation is inherently a nonlinear optical effect.
In our work, we assume that the electric field is sufficiently small such that perturbation theory works (we discuss in Sec.~\ref{sec:discussion} how small it should be).
While even-order responses to $E$ vanishes in centrosymmetric systems, they are allowed in noncentrosymmetric systems.
The bulk photovoltaic effect studied in the present paper is thus primarily a second-order response.

The second-order optical response under the uniform illumination of light has the form $j^c(\omega_1+\omega_2)=\sigma^{c;ab}(\omega_1+\omega_2;\omega_1,\omega_2)E_a(\omega_1)E_b(\omega_2)$ in general.
Let us focus on the DC (direct current) generation
\begin{align}
\label{eq:second-order_dc}
j^c_{\rm DC}
&=\sigma^{c;ab}_{\rm DC}(\omega)E_a(\omega)E_b(-\omega),
\end{align}
where $j^c_{\rm DC}=j^c(0)$, and $\sigma^{c;ab}_{\rm DC}(\omega)=\sigma^{c;ab}(0;\omega,-\omega)$.
In the clean limit --- where the interband relaxation rate is smaller than the photon frequency and the band gap, interband transitions are described by two processes: shift and injection~\footnote{Recently, another process has been proposed for metallic systems, which was termed {\it resonant photovoltaic effect}~\cite{bhalla2020resonant}.
We do not consider this effect here.}.
\begin{align}
\sigma^{c;ab}_{\rm DC}
=
\sigma^{c;ab}_{\rm shift}
+\sigma^{c;ab}_{\rm inj}.
\end{align}
The shift and injection currents correspond to the current generated by the change of the electron position and velocity, respectively, during the interband transition of electrons~\cite{sipe2000second}.
One can see this by noting that the shift and injection conductivities have the form of the Fermi Golden rule~\cite{sipe2000second} (See Appendix~\ref{sec:Golden} for explicit calculations).
Explicitly, they have the form~\cite{aversa1995nonlinear,de2020difference}
\begin{align}
\label{eq:shift-inj}
\sigma^{c;ab}_{\rm shift}
&=-
\frac{\pi e^3}{\hbar^2}
\int_{\bf k}
\sum_{n,m}
f^{\rm FD}_{nm}
(R^{c,a}_{mn}-R^{c,b}_{nm}) r^b_{nm}r^a_{mn}
\delta(\omega_{mn}-\omega)\notag\\
\sigma^{c;ab}_{\rm inj}
&=-
\tau \frac{2\pi e^3}{\hbar^2}
\int_{\bf k}
\sum_{n,m}
f^{\rm FD}_{nm}\Delta^c_{mn}r^b_{nm}r^a_{mn}\delta(\omega_{mn}-\omega),
\end{align}
where $\int_{\bf k}=\int d^dk/(2\pi)^d$, $f^{\rm FD}_{n}$ is the Fermi-Dirac distribution of the band $n$, $f^{\rm FD}_{nm}=f^{\rm FD}_{n}-f^{\rm FD}_{m}$, $\hbar\omega_{mn}=\hbar\omega_m-\hbar\omega_n$ is the energy difference between bands $m$ and $n$, $H\ket{n}=\hbar\omega_n\ket{n}$, $r^a_{mn}=\braket{m|i\d_a|n}$ and $v^c_{mn}=\hbar^{-1}\braket{m|\d_cH|n}$, and we use the notation $\d_a=\d/\d k_a$.
$R^{c;a}_{mn}=r^c_{mm}-r^c_{nn}+i\d_c\log r^a_{mn}$ is called the shift vector --- characterizing the interband-transition of the displacement, and $\Delta^c_{mn}=v^c_{mm}-v^c_{nn}$ is the interband-transition of the velocity.
$\tau$ is the relaxation time that saturates the injection current: withtout it, the injection of moving electrons and holes leads to a constant growth in time.
We take the electron charge as $-e$ (i.e., $e>0$).
Eq.~\eqref{eq:shift-inj} is valid with and without time-reversal symmetry.

\section{Symmetry and Power-Counting Analysis}
\label{sec:symmetry}

To get a general perspective on the low-energy properties of the shift and injection currents, we review their symmetry properties and then study the pattern of low-frequency divergences.
The key properties presented in this section are summarized in Table~\ref{tab:symmetry}, and they serve as basic ingredients for the analysis in Sec.~\ref{sec:model}.

\subsection{Symmetry of the shift and injection conductivities}

Let us first decompose the second-order DC conductivity into its real and imaginary parts:
\begin{align}
\sigma^{c;ab}(\omega)
=\sigma_L^{c;ab}(\omega)+i\sigma_C^{c;ab}(\omega).
\end{align}
Using $E^*(\omega)=E(-\omega)$ and Eq.~\eqref{eq:second-order_dc}, one can see that the conductivity can be symmetrized such that
\begin{align}
\label{eq:ab-parity}
\sigma_L^{c;ab}(\omega)
&=\sigma_L^{c;ba}(\omega),\notag\\
\sigma_C^{c;ab}(\omega)
&=-\sigma_C^{c;ba}(\omega).
\end{align}
Therefore, we always consider conductivity tensors satisfying Eq.~\eqref{eq:ab-parity}.
We note that the expressions in Eq.~\eqref{eq:shift-inj} are already symmetrized.
Assuming the form ${\bf E}(t)=|E|e^{-i\omega t}(\cos\phi,\sin\phi,0)+c.c.$ for linearly polarized light and ${\bf E}(t)=|E|e^{-i\omega t}(1,i,0)+c.c.$ for circularly polarized light, the real part of the second-order optical conductivity is
\begin{align}
\label{eq:currents}
{\rm Re}j^c_{\rm L-dc}
=(
&
\sigma^{c;xx}_L\cos^2\phi
+\sigma^{c;yy}_L\sin^2\phi\notag\\
&+2\sigma^{c;xy}_L\sin\phi\cos\phi)|E^2|\notag\\
{\rm Re}j^c_{\rm C-dc}
=(
&\sigma^{c;xx}_L
+\sigma^{c;yy}_L
-2\sigma^{c;xy}_C)|E^2|.
\end{align}
The real part of the conductivity is responsible for the current generation regardless of the polarization,while the imaginary part of the conductivity is responsible for the current generated by the circularly polarized light.
If one measures the current difference between the ones generated by left-circularly polarized light and right-circularly polarized light $\propto \sigma^{c;xy}-\sigma^{c;yx}$, only the imaginary part contributes.
In the following, we call the real part $\sigma^{c;ab}_L$ as linear conductivity and the imaginary part $\sigma^{c;ab}_C$ as circular conductivity.

From the definition in Eq.~\eqref{eq:second-order_dc} and the transformation properties of the current and electric field, it is clear that the second-order optical conductivity transforms like a third-rank tensor under spatial transformations.
That is, $\sigma'^{c_1;a_1b_1}_{\rm DC}={\cal M}_{c_1c}{\cal M}_{a_1a}{\cal M}_{b_1b}\sigma^{c;ab}_{\rm DC}$ under a point-group symmetry transformation $x_a\rightarrow x'_a={\cal M}_{ab}x_b$~\footnote{
We define $\sigma'^{c;ab}_{\rm DC}$ by $j'^c_{\rm DC}=\sigma'^{c;ab}_{\rm DC}E'_aE'_b$, where
$j'^{c_1}={\cal M}_{c_1c}j^{c}$ and $E'_{a_1}={\cal M}_{a_1a}E_{a_2}$.
}.
However, one should be careful when taking time reversal for relaxational processes as is well-known from the Onsager reciprocity relations in linear response theory~\cite{onsager1931reciprocal,kubo1957statistical,tokura2018nonreciprocal}.
For example, it seems like Eq.~\eqref{eq:second-order_dc} implies that time reversal reverses the sign of the DC conductivity for linearly polarized light.
However, we need to additionally reverse the sign of phenomenological relaxation rate $\Gamma$, in order to make the decay in time to the growth in time.
The correct time reversal for the second-order DC conductivity is
\begin{align}
\sigma^{c;ab}_{\rm DC}(\omega,\Gamma)
\rightarrow
\sigma'^{c;ab}_{\rm DC}(\omega,\Gamma)
=-\sigma^{c;ba}_{\rm DC}(\omega,-\Gamma).
\end{align}
See Appendix~\ref{sec:time}.
When applying this equality, the delta function and $\tau$ should be interpreted as $\pi^{-1}\Gamma/[(\omega_{mn}-\omega)^2+\Gamma^2]$ and $\Gamma^{-1}$, so they reverses sign under $\Gamma\rightarrow -\Gamma$.
Thus, the real part of the shift and injection conductivity tensor transforms as
\begin{align}
\label{eq:linear-transform}
\sigma'^{c_1;a_1b_1}_{\rm shift,L}
&=
{\cal M}_{c_1c}{\cal M}_{a_1a}{\cal M}_{b_1b}\sigma^{c;ab}_{\rm shift,L}\notag\\
\sigma'^{c_1;a_1b_1}_{\rm inj,L}
&=(-1)^{s_T}
{\cal M}_{c_1c}{\cal M}_{a_1a}{\cal M}_{b_1b}\sigma^{c;ab}_{\rm inj,L},
\end{align}
and the imaginary part transforms as
\begin{align}
\label{eq:circular-transform}
\sigma'^{c_1;a_1b_1}_{\rm shift,C}
&=(-1)^{s_T}
{\cal M}_{c_1c}{\cal M}_{a_1a}{\cal M}_{b_1b}\sigma^{c;ab}_{\rm shift,C}\notag\\
\sigma'^{c_1;a_1b_1}_{\rm inj,C}
&=
{\cal M}_{c_1c}{\cal M}_{a_1a}{\cal M}_{b_1b}\sigma^{c;ab}_{\rm inj,C}.
\end{align}
under the spacetime symmetry transformation $(t,x_a)\rightarrow (t',x_a')=((-1)^{s_T}t,{\cal M}_{ab}x_b)$.
Alternatively, one can verify these transformation rule by examining transformations of $R^{c,a}_{mn}$, $r^a_{nm}$, $\Delta^c_{mn}$ from the form in Eq.~\eqref{eq:shift-inj}.
See Appendix~\ref{sec:transformation} for a derivation.

Knowing these symmetry transformation properties, one can use the MTENSOR~\cite{gallego2019automatic} in the Bilbao Crystallographic Server to see which tensor components are required to vanish by symmetry for any of 80 magnetic point groups.
For our purpose, the most important symmetries are simply time reversal $T$ and spacetime inversion $PT$ symmetries (note again that $P$ is always broken in the present paper).
We summarize their role in Table.~\ref{tab:symmetry}.
While only linear shift and circular injection currents can be generated in time-reversal-symmetric systems, they vanish in $PT$-symmetric systems, so linear injection and circular shift currents can be generated.
time-reversal symmetry and spacetime inversion symmetry are thus complementary to each other, as pointed out in~\cite{holder2020consequences}.
This behavior is manifested in the geometric quantities related to the responses.
As we show below in Sec.~\ref{sec:geometry}, $T$-symmetric responses are related to both of the Berry curvature and the quantum metric, while $PT$-symmetric responses are related to the quantum metric only.
In general magnetic systems without $T$ and $PT$ symmetries, all of the four types of currents can be generated.

When there is $MT$ or $C_2T$ symmetry instead, where $M$ and $C_2$ are mirror and twofold rotation, a phenomenon of "separation of responses" occurs, meaning that different directions manifest different types of responses.
It is because those symmetries act like time reversal in some directions and act like spacetime inversion in the other directions.
For example, $M_xT$ acts like time reversal in the $y$ and $z$ directions, whereas it acts as spacetime inversion in the $x$ direction.
In this case, $x$-polarized light generates a shift current along $y$ and $z$ while generating an injection current along $x$.
We demonstrate the separation of responses through model calculations in Sec.~\ref{sec:model} and also through first-principles calculations in Sec.~\ref{sec:first-principles}.

\subsection{Power-counting analysis of the low-energy divergence}

Let us now examine the low-energy divergence of the second-order responses in semimetals.
We can estimate the divergence by counting the power of photon frequency in Eq.~\eqref{eq:shift-inj}.
Since the delta function has dimension $\omega^{-1}$, $R^c$ and $r^a_{mn}$ has dimension $k^{-1}$, and $\Delta^c$ has dimension $\omega/k$, the shift and injection conductivity scales as
\begin{align}
\label{eq:scale-k}
\sigma^{c;ab}_{\rm shift}
\sim \frac{e^3}{\hbar^2}\frac{1}{\omega}k^{d-3},\notag\\
\sigma^{c;ab}_{\rm inj}
\sim \frac{e^3}{\hbar^2}\tau k^{d-3}.
\end{align}
where $k$ is the characteristic wave vector.
When the system has dispersion $E\propto k^{\alpha}$, $k\sim E^{1/\alpha}\sim \omega^{1/\alpha}$.
Thus, smaller $\alpha$ is preferred to get large optical responses for small $\omega$ in 2D, while it is independent on the dispersion in 3D.

In lattice systems, stable Weyl points always (and also Dirac points protected by symmorphic symmetries~\cite{yang2014classification}) appear pairwise according to the Nielsen-Ninomiya theorem~\cite{nielsen1981no,witten2015three}.
Therefore, a response in semimetals should be a sum of responses from different Weyl or Dirac points.
However, different gap-closing points are located at different energy levels in general, so it is expected that only a particular point contributes to the low-energy response significantly~\cite{de2017quantized}.
An exact cancellation or reinforcement among different gap-closing points can occur due to symmetries, but it can be considered straightforwardly from the symmetry transformation properties of the conductivity tensor.
In this regard, the response from a single gap-closing point can be associated with the low-energy response of a whole system.

Thus, we consider a gap-closing point with a linear dispersion, described by a Dirac Hamiltonian
\begin{align}
\label{eq:Dirac}
H_0({\bf k})=\hbar v\sum_{a=1}^dk_a\Gamma_a,
\end{align}
where $\Gamma_a$ are mutually anticommuting matrices, such that the spectrum has the form
\begin{align}
E({\bf k})
=\pm \hbar v|{\bf k}|.
\end{align}
In this case, we have
\begin{align}
\sigma^{c;ab}_{\rm shift}
&\sim \frac{e^3}{\hbar^2}\frac{1}{\omega}\left(\frac{v}{\omega}\right)^{3-d},\notag\\
\sigma^{c;ab}_{\rm inj}
&\sim \frac{e^3}{\hbar^2}\tau \left(\frac{v}{\omega}\right)^{3-d}
\end{align}
through a dimensional analysis.
This divergence is expected to occur in the absence of a symmetry cancellation.
However, Eq.~\eqref{eq:Dirac} has too many symmetries, so we need to break them in general.
A Dirac point has inversion symmetry (by definition it is nonchiral), so the second-order optical response is forbidden.
Also, as noted in Ref.~\cite{chan2017photocurrents,yang2017divergent,holder2020consequences}, a Weyl point described by Eq.~\eqref{eq:Dirac} does not show a second-order optical response by linearly polarized light.
It is because Eq.~\eqref{eq:Dirac} has SO($d$) rotational symmetry in $d$ spatial dimensions.
In 3D, the absence of mirror symmetry (chirality) in Weyl semimetals allows one  unique nonvanishing independent component $\sigma^{3;12}_{{\rm inj},C}$ under circularly polarized light.
Because a Weyl point in 3D has time-reversal symmetry around the gap-closing point, the circular shift current is also forbidden, and the generated DC current is the circular injection current.
Even for a more general linear dispersion described by $H=\sum_{a,i=1}^d\hbar v_{ai}k_a\Gamma_i$, which has apparently less symmetry, only circular injection currents for a Weyl point can be nonvanishing.
It is because the conductivity for this Hamiltonian is given by $\sigma^{c;ab}_{\rm DC}=v^{-3}v_{ai}v_{bj}v_{ck}\det(v_{ai}/v)\sigma^{k;ij}_{\rm DC,0}$~\cite{yang2017divergent}, where $\sigma^{k;ij}_{\rm DC,0}$ is the conductivity for $v_{ai}=v\delta_{ai}$, which is the case for Eq.~\eqref{eq:Dirac} (See Appendix~\ref{sec:k-linear} for a derivation).

\begin{figure}[t!]
\includegraphics[width=8.5cm]{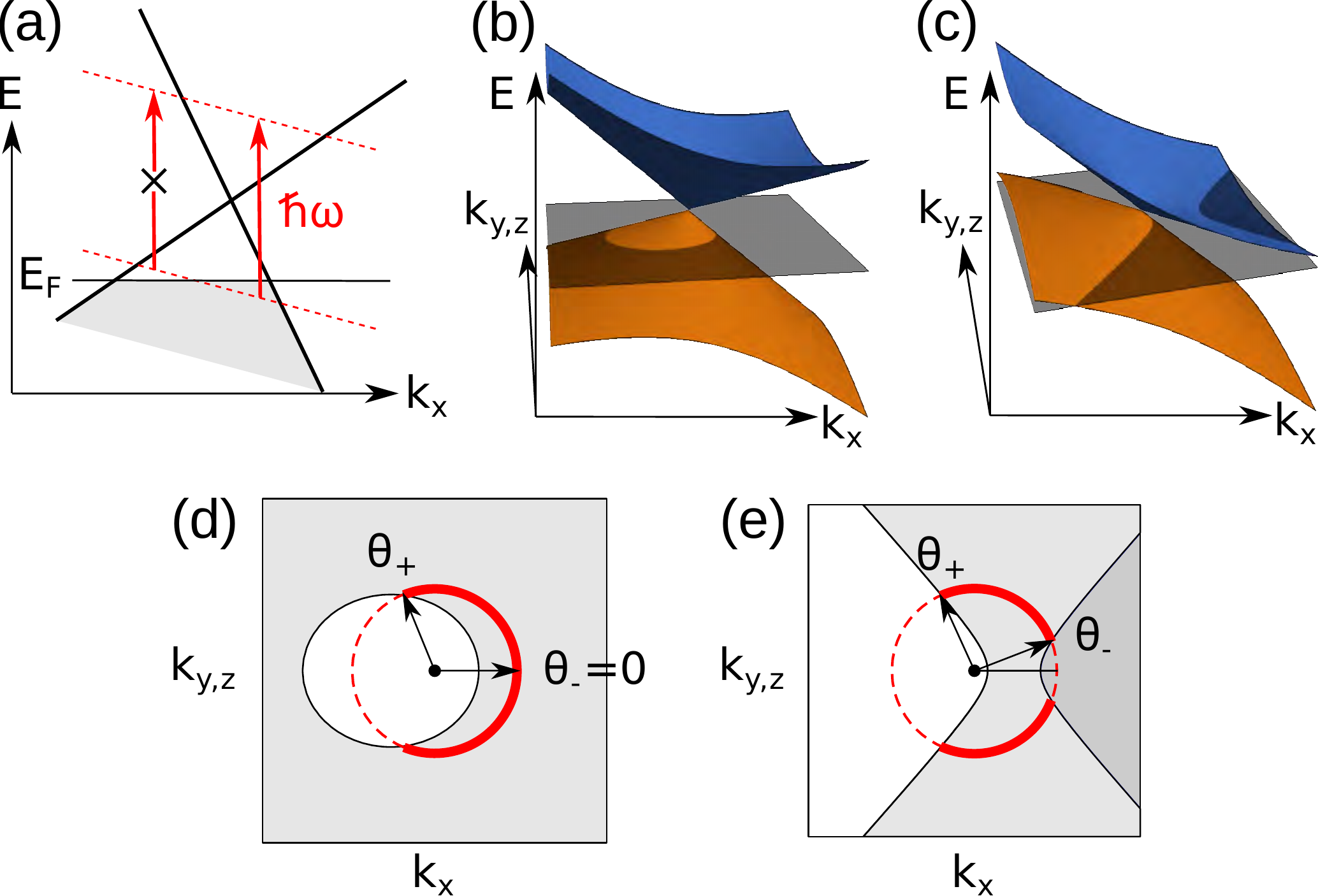}
\caption{Optical excitation near a tilted Dirac or Weyl cone.
(a) Excitation of electrons by the absortion of a photon with requency $\omega$.
(b) Type-I spectrum. $|v'/v|<1$.
(c) Type-II spectrum. $|v'/v|>1$.
Here, $v$ is the velocity of a Dirac or Weyl fermion at zero tilting, and $v'$ is the overall velocity shift by $\mu\rightarrow \mu+\hbar v'k_x$, giving rise to a tilting of the cone.
The gray planes in (b,c) show the Fermi level.
(d,e) Optically active region in momentum space for (d) type-I and (e) type-II cases. 
Small black dots at the center indicates the location of the gap-closing point.
Red circles around the point shows the surface satisfying $\hbar\omega=\hbar\omega_c-\hbar\omega_v=2\hbar v k$, where $k=0$ at the gap closing.
Both conduction and valence bands are unoccupied in the white region, only the valence band is occupied in the light gray region, and both bands are occupied in the gray region.
Electrons can be optically excited only on the solid red arcs (i.e., $\theta_-<\theta<\theta_+$, where $\theta$ is the absolute value of the polar angle in either 2D or 3D), which is in the light gray region.
}
\label{fig:tilted}
\end{figure}

The only way to generate the leading divergence for shift currents and linear injection currents is to tilt the Dirac or Weyl cone, as shown in Fig.~\ref{fig:tilted}.
Since it allows anisotropic optical excitations around the gap-closing point, photocurrents can flow whose direction depends on the direction of the tilting.
To see that other symmetry breaking gives subleading power in $\omega^{-1}$, let us add symmetry-breaking perturbations to the Dirac Hamiltonian.
\begin{align}
H({\bf k})
=H_0({\bf k})+\hbar\sum_a\sum_{p=0}^{\infty}\lambda_{p,a}k^p\Gamma_{a}.
\end{align}
Here, $\lambda_{p,a}$ is a constant parameter, $\Gamma_{a\ne 0}$ are mutually anticommuting matrices, and we also include $\Gamma_0$ as the identity matrix.
Since the dimensionless parameter is $\lambda_{p,a}k^p\omega^{-1}\sim \lambda_{p,a}\omega^{p-1}$, responses due to perturbations in $\lambda_{p,a}$ have weaker low-frequency divergences for $p>1$.
For example, let us consider a quadratic correction, resulting in the dispersion relation $E=\pm(\hbar vk+\lambda_2k^2 )$.
Optical excitations occur in the region satisfying $\omega=E_+-E_-=2(\hbar vk+\lambda_2k^2 )$, i.e. $k=\omega/v (1-\lambda_2\omega/v^2 )+O(\omega^3 )$. By inserting this to Eq.~\eqref{eq:scale-k}, we have corrections to the leading divergence by a fraction $\left|(d-3)\lambda_2\omega/v^2\right|\ll1$ for small $\omega$.
At the $p=0$ order, $\lambda_{0,a}$ comes as either the chemical potential, the shift of the location of the gap-closing point, or a mass term~\footnote{If we allow transforming a $4\times 4$ Dirac point to different kinds of nodes, other possibilities exist, which includes the splitting of a Dirac point into two Weyl points or the inflating of a Dirac point to a nodal line.
These reqruie $m_{ab}\Gamma_a\Gamma_b$ terms.
In the former case each of the Weyl points can be understood in our framework.
Understanding the second-order response of nodal lines appearing in the latter case is an interesting topic, but it is out of the scope of this work.}, i.e., we can write the Hamiltonian as $H({\bf k})=-\mu+\sum_{a=1}^d(k-k_0)_a\Gamma_a+M\sum\Gamma_{a>d}$.
None of them generate nonvanishing shift and linear injection currents.
$\mu$ breaks no symmetry, ${\bf k}-{\bf k}_0$ can be redefined as ${\bf k}$ such that gap closes at ${\bf k}=0$, and the mass term can serve as an inversion symmetry operator so that it forbids second-order optical responses.
Let us now consider $p=1$.
As we show above, the linear spectrum without tilting has zero shift and linear injection current responses.
Thus, the remaining possibility is tilting the cone by adding $\lambda_{1,0}=-v'\cos\theta$: this gives an overall tilting of energy levels by $\mu\rightarrow \mu+v'k_x$.

Our analysis shows that the shift and injection conductivity tensors of tilted massless Dirac and Weyl points have the form
\begin{align}
\label{eq:leading-divergence}
\sigma^{c;ab}_{\rm shift}
&=\frac{e^3}{\hbar^2}\frac{1}{\omega}\left(\frac{v}{\omega}\right)^{3-d}
{\cal F}^{c;ab}_{\rm shift}\left(\frac{2\mu}{\hbar\omega},\frac{v'}{v}\right),\notag\\
\sigma^{c;ab}_{\rm inj}
&=\frac{e^3}{\hbar^2}\tau \left(\frac{v}{\omega}\right)^{3-d}
{\cal F}^{c;ab}_{\rm inj}\left(\frac{2\mu}{\hbar\omega},\frac{v'}{v}\right),
\end{align}
Let us explain how ${\cal F}$s depend on $v'/v$ and $2\mu/\hbar\omega$ in general.
When $\mu\ne 0$, i.e., when the Fermi level is away from the gap-closing point, the chemical potential sets the lowerbound for frequency, so ${\cal F}$s do not diverge as $\omega\rightarrow 0$.

When the Fermi level is exactly at the gap-closing point, i.e., $\mu=0$, ${\cal F}$s show significantly different behaviors for $|v'/v|<1$ and $|v'/v|>1$, which are called type-I and type-II~\cite{soluyanov2015type}, respectively [See Fig.~\ref{fig:tilted}(b,c)].
In the type-I case, tilting cannot generate shift and injection currents when $\mu=0$.
It is because, in this case, the Fermi surface is a point, so anisotropic excitation cannot occur.
Thus, only circular injection currents can be generated, which do not need tilting for its generation.

In contrast, in the type-II case, the Fermi surface has a finite size at $\mu=0$, so anisotropic excitation can occur in principle.
However, it depends on whether the response is $T$-symmetric or $PT$-symmetric.
While the $T$-symmetric responses (linear shift and circular injection) has a nontrivial response at $\mu=0$~\cite{yang2017divergent}, the $PT$-symmetric responses (linear injection and circular shift) has a vanishing response at $\mu=0$.
This is related to the fact that a Hamiltonian with only ${\bf k}$-linear terms has an emergent $CPT$ symmetry, where $C$ is the particle-hole operator, and $CPT=1$:
$(CPT)H({\bf k})(CPT)^{-1}=-H({\bf k})$.
At $\mu=0$, $PT$-symmetric responses should be zero since they respect an effective $C$ symmetry also (where $(PT)^{-1}$ takes the role of an effective $C$ operator)
 --- which reverses the direction of the current, while $T$-symmetric responses lack $C$ symmetry such that they can be nontrivial.

This shows that magnetic and nonmagnetic systems have the same low-frequency divergent behavior when $\mu=0$ exactly.
Nevertheless, as far as $\mu$ is small but not exactly zero, we can still expect enhanced $PT$-symmetric responses at small frequency $\omega\sim 2\mu/\hbar$ by a factor in front of ${\cal F}$s in Eq.~\eqref{eq:leading-divergence}.
Therefore, $PT$-symmetric responses in magnetic systems also can show a divergent behavior associated with the $\omega\rightarrow 0$ limit with a fixed ratio of $2\mu/\hbar\omega$.

We investigate the symmetry properties of the tilted Dirac and Weyl points more closely in Sec.~\ref{sec:model} by explicitly calculating the conductivity tensors.
Before that, we derive the general formula for the shift and injection conductivities for arbitrary Dirac Hamiltonians and provide their geometric aspects in the following sections.
It adds more perspective on the transformation rule Eqs.~\eqref{eq:linear-transform} and~\eqref{eq:circular-transform} and the divergent behavior near the gap-closing point.

\section{Quantum geometric aspects}
\label{sec:geometry}

In the previous section, we analyze the overall trend of shift and injection currents using symmetry and power-counting analysis.
Here we show that every detail of the conductivity profile for the shift and injection currents is determined by quantum geometric quantities in the low-frequency regime.
It was pointed out in Refs.~\cite{morimoto2016topological,nagaosa2017concept} that the shift current is related to quantum geometry because the shift vector includes the Berry connection --- a geometric quantity.
Following these works, the geometric nature of the shift and injection currents was previously discussed in several works~\cite{de2017quantized,holder2020consequences}. However, no simple quantitative relationship between the response and the geometric quantities has been found except for the circular injection conductivity~\cite{de2017quantized}.
In this section, we show that the shift and injection conductivities are proportional to geometric quantities that have natural geometric meaning on the Bloch sphere.
These relationships are not limited to massless Dirac and Weyl points and are exact for any two-band system or $PT$-symmetric four-band system.
It implies that interband-transitive photovoltaic responses at the low-frequency regime probe the quantum geometry of materials.
In this perspective, the low-frequency divergence of the shift and injection current responses of gap-closing points can be attributed to their geometrically singular nature.
Also, time reversal symmetry transformation of the conductivity tensors, which are quite confusing, can be simply understood from the transformation properties of the geometric quantities.
In the following, we first derive the formula that relates injection and shift currents with the Bloch vector of general Dirac Hamiltonians with arbitrary matrix size in any spatial dimensions.
Then, we provide a quantum geometric interpretation of our formula.

\subsection{Shift and injection conductivity for Dirac Hamiltonians}

We consider the low-energy effective model systems described by the following $d_M\times d_M$ Dirac Hamiltonian
\begin{align}
\label{eq:general-Dirac}
H({\bf k})=-\mu({\bf k})+\sum_{i}f_i({\bf k})\Gamma_i,
\end{align}
where $\Gamma_i$ are mutually anticommuting matrices.
This Hamiltonian described a single Dirac or Weyl point when $f_i({\bf k})=k_i$, but here we do not need to assume linear dispersion and consider general form of $f_i({\bf k})$s.
In particular, the above Hamiltonian describes general two-band Hamiltonians when $d_M=2$, where $\Gamma_{i=1,2,3}$ are three Pauli matrices, and it describes general four-band $PT$-symmetric Hamiltonians (with $(PT)^2=-1$) when $d_M=4$, where $\Gamma_{i=1,\hdots,5}$ are five Gamma matrices.

Let us express the injection and shift conductivities in terms of $f_i$s.
This makes theoretical analysis and numerical calculations convenient.
As for the injection current, one can integrate the delta function easily by using $\Delta^c_{mn}=\d_c\omega_{mn}$ such that $\Delta^c_{mn}\delta(\omega_{mn}-\omega)=\d_c\Theta(\omega_{mn}-\omega)$.
After that, we obtain
\begin{align}
\label{eq:inj-geometry}
\sigma^{c;ab}_{\rm inj}
&=
-\tau \frac{2\pi e^3}{\hbar^2}
\int_{\omega_{cv}=\omega}
\frac{d^{d-1}k}{(2\pi)^d}(\hat{n}\cdot\hat{c})
Q_{ba}
\end{align}
for $\omega>0$, where $\hbar\omega_{cv}$ is the energy gap between the conduction and valence bands, $\hat{n}$ is the surface normal vector, and
\begin{align}
\label{eq:QGT}
Q_{ba}
&=
\sum_{n\in \rm occ}\sum_{m\in\rm unocc}r^b_{nm}r^a_{mn}\notag\\
&=\sum_{i,j}\d_bf_i\d_af_j\frac{d_M(\delta_{ij}-\hat{f}_i\hat{f}_j+iJ_{ij})}{8f^2}
\end{align}
is the so-called quantum geometric tensor~\cite{provost1980riemannian} (See Appendix~\ref{sec:QGT} for a derivation of the second equality).
Here, $f=\sqrt{\sum_{i=1}f_i^2}$, and $J_{ij}=-i\sum_k\epsilon_{ijk}\hat{f}_k$ for $d_{M}=2$ and $J_{ij}=0$ for $d_{M}=4$.
The vanishing of $J_{ij}$ for $d_M=4$ is due to the presence of $PT$ symmetry.
$Q_{ba}$ is called the quantum geometric tensor because its real and imaginary parts are related to the quantum metric $g_{ba}$ and the Berry curvature $F_{ba}$ by
\begin{align}
Q_{ba}
=g_{ba}-\frac{i}{2}F_{ba}.
\end{align}
The relationship between the circular injection current and the Berry curvature was found in Ref.~\cite{de2017quantized}.
On the other hand, the role of the quantum metric in determining the linear injection current was not discussed in the literature.
We explain more on the geometric meaning of the quantum geometric tensor in Sec.~\ref{sec:geometric-meaning}.

The shift current has more complicated form, and it can be related to the matrix elements of the derivatives of the Hamiltonian as~\cite{cook2017design}
\begin{align}
\label{eq:shift-matrix}
&R^{c,a}_{mn}r^b_{nm}r^a_{mn}
=
i\frac{ v^b_{nm}}{\omega_{mn}^2}
\left[
w^{ac}_{mn}
-\frac{
 v^c_{mn}\Delta^a_{mn}
+ v^a_{mn}\Delta^c_{mn}}
{\omega_{mn}}
\right]\notag\\
&+\frac{v^b_{nm}}{\omega^2_{mn}}
\sum_{\omega_p\ne \omega_m,\omega_n}\left(\frac{v^c_{mp}v^a_{pn}}{\omega_{mp}}-\frac{v^a_{mp}v^c_{pn}}{\omega_{pn}}\right),
\end{align}
where $w^{ac}_{mn}=\hbar^{-1}\braket{m|\d_a\d_cH|n}$ is the diamagnetic term.
The second line involves virtual transitions among three different bands, so it vanishes in our Dirac system that has only two energy levels with energy $\hbar\omega_c=f_0+f$ and $\hbar\omega_v=f_0-f$, where $f=\sqrt{\sum_{i=1}f_i^2}$.
It then follows that~\footnote{
In fact, the integrand should be $(C_{bca}-C_{acb}^*)/2$ since the integrand has the form $(R^{c,a}_{mn}-R^{c,b}_{nm})r^b_{nm}r^a_{mn}$, but $C_{bca}$ gives the same value of the real part of the current.
For linearly polarized light, the conductivity is symmetric under $a\leftrightarrow b$. In this case, $C_{bca}+C^*_{acb}+a\leftrightarrow b$ is real, such that it contributes to the imaginary part of the conductivity, and thus to the imaginary part of the current.
For circularly polarized light, since $C_{bca}+C^*_{acb}-a\leftrightarrow b$ is imaginary, the conductivity takes a real value, so it contributes to the imaginary part of the current for circularly polarized light.}
\begin{align}
\label{eq:shift-geometry}
\sigma^{c;ab}_{\rm shift}(\omega)
&=
-i\frac{2\pi e^3}{\hbar^2}
\int_{\bf k}
f^{\rm FD}_{vc}
C_{bca}
\delta(\omega_{cv}-\omega)
\end{align}
for $\omega>0$, where
\begin{align}
\label{eq:shift-Dirac}
&C_{bca}
=-i\sum_{n\in \rm occ}\sum_{m\in \rm unocc}R^{c,a}_{mn}r^b_{nm}r^a_{mn}\notag\\
&\qquad =\frac{d_M}{8f^2}
\sum_{i,j}
(\delta_{ij}-\hat{f}_i\hat{f}_j+iJ_{ij})\times\notag\\
&
\bigg[
\d_bf^i\d_a\d_cf^j-\frac{1}{f}
\left(
\d_bf^i\d_cf^j\d_af
+\d_bf^i\d_af^j\d_cf
\right)
\bigg].
\end{align}
A special case of this formula was derived in Ref.~\cite{cook2017design} for two-band models with time-reversal symmetry.
Our formula in Eq.~\eqref{eq:shift-Dirac} extends the existing formula to describe arbitrary systems described by the Dirac Hamiltonian in Eq.~\eqref{eq:general-Dirac}.
We show below that $C_{bca}$ has geometric meaning as a connection.

\subsection{Geometry on the generalized Bloch sphere}
\label{sec:geometric-meaning}

Let us explain the geometric meaning of the quantum geometric tensor $Q_{ba}$ as geometric quantities defined on the generalized Bloch sphere.
This point of view is helpful for understanding the geometric meaning of $C_{bca}$ as well as that of $Q_{ba}$.

We consider the following general Dirac Hamiltonian in Eq.~\eqref{eq:general-Dirac}.
Then, the generalized Bloch vector ${\bf f}({\bf k})$ is a map
\begin{align}
{\bf f}:{\rm BZ}\rightarrow {\bb R}^{d_{\Gamma}},
\end{align}
where $d_{\Gamma}$ is the number of Gamma matrices.
This map defines a pull-back of the quantum geometric tensor from the ${\bf f}$-space to the Brillouin zone.

Let us recall that the quantum geometric tensor has the following form
\begin{align}
Q_{ab}
&=\sum_{i,j}\d_af_i\d_bf_jq_{ij},
\end{align}
where
\begin{align}
q_{ij}
=\frac{d_M(\delta_{ij}-\hat{f}_i\hat{f}_j+iJ_{ij})}{8f^2},
\end{align}
and $f=|{\bf f}|$.
This is a pull-back of the quantum geometric tensor $q_{ij}$ defined in the ${\bf f}$-space to the momentum space by a transformation $\d_af_i$ of tangent vectors
\begin{align}
\d_{a}=(\d_af_i) \d_i,
\end{align}
where $\d_i=\d/\d f_i$.
The quantum metric and the Berry curvature are given by
\begin{align}
g_{ab}
&= \sum_{i,j}\d_af_i\d_bf_j\eta_{ij},\notag\\
\frac{1}{2}F_{ab}
&= \sum_{i,j}\d_af_i\d_bf_j\epsilon_{ij},
\end{align}
where $\eta_{ij}$ and $\epsilon_{ij}$ are the real and imaginary parts of $q_{ij}$, i.e., $q_{ij}=\eta_{ij}-i\epsilon_{ij}$.
In this viewpoint, the quantum metric and the Berry curvature are pull-backs of the metric $\eta_{ij}$ and the symplectic form $\epsilon_{ij}$ defined on the Bloch sphere.
The metric $\eta_{ij}$ measures length $ds$ through $ds^2=\eta_{ij}df^idf^j$, and the symplectic form $\epsilon_{ij}$ measures the oriented area $dA$ through $dA=\epsilon_{ij}dx^idx^j$ in the ${\bf f}$-space.

While the geometric quantity $\eta_{ij}$ is defined on the whole ${\bf f}$ space, it can be regarded to measure the length on the unit sphere with $f=1$, and it is irrelevant for the radial direction $f=|{\bf f}|$~\footnote{A metric having this property is called the Fubini-Study metric, so the quantum metric is often called as the Fubini-Study metric~\cite{kolodrubetz2017geometry}.}.
For example, $ds^2=(1/4)(d\theta^2+\sin^2\theta d\phi^2)$ in polar coordinates when $d_M=2$.
To see this without a coordinate transformation, first note that $P_{ij}=\delta_{ij}-\hat{f}_i\hat{f}_j$ is the projection to the plane perpendicular to $\hat{f}$.
The metric measures the length only along the angular directions on a sphere with a fixed $f$.
Also, the $f^{-2}$ factor normalizes the length such that only the angle between two points on a sphere is measured.
Similarly, the symplectic form also measures the area on the unit sphere.
In this sense, the quantum geometric tensor is a geometric quantity defined on the generalized Bloch sphere ($f=1$).

Another geometric quantity called the Levi-Civita connection also can be constructed on the generalized Bloch sphere.
Its components are called the Christoffel symbols, and they can be written in two ways --- the first and the second kind.
The Christoffel symbols of the second kind $\gamma^{k}_{ij}$ are defined by
\begin{align}
\label{eq:second-kind}
\d_i{\bf e}_j
=\sum_k\gamma^k_{ij}{\bf e}_k,
\end{align}
where ${\bf e}_i$ is the projection of the unit vector along the $f_i$ direction to the tangent space by $P_{ij}=\delta_{ij}-\hat{f}_i\hat{f}_j$.
It measures how vectors and tensors are changed as we move them parallel to the direction of the curved surface (which is the generalized Bloch sphere in our case)~\footnote{This property is shared by another connection, the nonabelian Berry connection $A^a_{mn}=\braket{m|i\d_a|n}$, but the difference is that $\gamma^k_{ij}$ is the connection for the Bloch vector ${\bf f}$ while $A^a_{mn}$ is the connection for the quantum state $\ket{n}$.}.
We have
\begin{align}
\gamma^{k}_{ij}
&=-\frac{f_i}{f^2}(\delta_{jk}-\hat{f}_j\hat{f}_k)-\frac{f_j}{f^2}(\delta_{ik}-\hat{f}_i\hat{f}_k).
\end{align}
It is related to the metric tensor $\eta_{ij}$ by
\begin{align}
\gamma^{k}_{ij}
&\equiv \sum_l(\eta^{-1})^{kl}\frac{1}{2}\left(\d_i\eta_{jl}+\d_{j}\eta_{il}-\d_l\eta_{ij}\right),
\end{align}
where $(\eta^{-1}\eta)_{ij}=\delta_{ij}-\hat{f}_i\hat{f}_j$.
Using the Christoffel symbols of the second kind and the quantum geometric tensor, we define the Christoffel symbols of the first kind as
\begin{align}
\gamma_{kij}
&\equiv \sum_l\eta_{kl}\gamma^{l}_{ij}\notag\\
&=-\frac{f_i}{f^2}\eta_{kj}-\frac{f_j}{f^2}\eta_{ki}.
\end{align}
Here we distinguish the first and second Christoffel symbols by the uppercase and lowercase letters for the first component, while we do not distinguish the uppercase and lowercase for other quantities.
We can also define a similar quantity using the symplectic form rather than the metric tensor by~\cite{gelfand1997fedosov}
\begin{align}
\label{eq:symplectic-symbol}
\tilde{\gamma}_{kij}
&\equiv \sum_l\epsilon_{kl}\gamma^{l}_{ij}\notag\\
&=-\frac{f_i}{f^2}\epsilon_{kj}-\frac{f_j}{f^2}\epsilon_{ki}.
\end{align}
To write $\gamma_{kij}$ and $\tilde{\gamma}_{kij}$ in a unified way, we introduce \begin{align}
c_{kij}
&=\gamma_{kij}-i\tilde{\gamma}_{kij}.
\end{align}
We call $c_{kij}$ as the {\it quantum geometric connection} in analogy with the quantum metric tensor.

\begin{table*}[t!]
\begin{tabular}{c|cccc|ccc}
System	&${\cal F}^{x;xx}_{{\rm inj},L}$	&${\cal F}^{x;yy}_{{\rm inj},L}$		&${\cal F}^{y;xy}_{{\rm inj},L}$		&${\cal F}^{y;xy}_{{\rm shift},C}$
&${\cal F}^{y;zx}_{{\rm shift},L}$		&${\cal F}^{x;yz}_{{\rm inj},C}$		&${\cal F}^{y;zx}_{{\rm inj},C}$
\\
\hhline{=|====|===}
3D	Weyl 
&$-\frac{1}{8}\cos^2\theta+\frac{1}{16}\cos^4\theta$	
&$-\frac{1}{16}\cos^2\theta-\frac{1}{32}\cos^4\theta$		
&$-\frac{1}{32}\sin^4\theta$
&$-\frac{1}{4}\sin\theta$
&$-\frac{1}{8}\cos\theta+\frac{1}{8}\cos^3\theta$	
&$-\frac{1}{12}\cos^3\theta$	
&$-\frac{1}{8}\cos\theta+\frac{1}{24}\cos^3\theta$\\
3D	Dirac 
&$-\frac{1}{4}\cos^2\theta+\frac{1}{8}\cos^4\theta$	
&$-\frac{1}{8}\cos^2\theta-\frac{1}{16}\cos^4\theta$		
&$-\frac{1}{16}\sin^4\theta$
&$-\frac{1}{2}\sin\theta$
&$0$	&$0$	&$0$\\
2D Dirac
&$\frac{1}{3}\sin^3\theta$	
&$\sin\theta-\frac{1}{3}\sin^3\theta$		
&$-\frac{1}{3}\sin^3\theta$
&$-\sin\theta$
&$0$	&$0$	&$0$
\end{tabular}
\caption{
${\cal F}^{c;ab}(\theta)$ of a single Weyl or Dirac point in two and three dimensions.
Here, $0\le \theta\le \pi$ is the absolute value of the polar angle from the $x$-axis.
Linear shift and circular injection parts vanishes for a Dirac point due to $PT$ symmetry.
}
\label{tab:tilted}
\end{table*}
The Levi-Civita connection does not transform like a tensor under coordinate transformations~\cite{nakahara2003geometry}, which is due to the derivative acting on tensorial quantities in the definition of the Christoffel symbols of the second kind [See Eq.~\eqref{eq:second-kind}].
The Christoffel symbols of the second kind defined on the generalized Bloch sphere $\gamma^{k}_{ij}$ are related to those defined in the Brillouin zone $\Gamma^{c}_{ab}$ by $\Gamma^{c}_{ab}
=\sum_d
(g^{-1})^{cd}\frac{1}{2}\left(\d_bg_{da}+\d_{a}g_{db}-\d_dg_{ab}\right)
=\sum_{i,j,l}
\d_lk^c\d_af_i\d_bf_j\gamma^{l}_{ij}
+\sum_{i,j}\d_ik^c\d_a\d_bf_j\delta_{ij},$
where the second term shows the non-tensorial transformation property.
It follows that the quantum geometric connection in the Brillouin zone has the form
\begin{align}
C_{cab}
&=
\sum_{i,j,k}
\d_cf_k\d_af_i\d_bf_jc_{kij}
+\sum_{ij}\d_cf_i\d_a\d_bf_jq_{ij},
\end{align}
where $C_{cab}=Q_{cd}\Gamma^{d}_{ab}$.
It is identical to the quantity defined in Eq.~\eqref{eq:shift-Dirac}, as one can see by using $\d_af=\sum_kf^{-1}f_k\d_af_k$.
This quantity, the quantum geometric connection, reveals the geometric nature of the low-frequency shift current in the most transparent way.
Let us note that, in general relativity, the equivalence principle requires that the Levi-Civita connection does not appear directly as an observable quantity, because it is not invariant under a coordinate transformation.
However, here we do not have such an equivalence principle for the Bloch vector ${\bf f}$, so it is allowed to observe the Levi-Civita connection (or quantum geometric connection).

\subsection{More on the geometric aspect of the shift current}

Equation~\eqref{eq:shift-geometry} shows that the linear (circular) shift current corresponds to the imaginary (real) part of the quantum geometric connection.
Thus, the circular shift current reveals the Christoffel symbol of the first kind $\Gamma_{cab}=\frac{1}{2}\left(\d_bg_{ca}+\d_{a}g_{cb}-\d_cg_{ab}\right)$.
On the other hand, the linear shift current is related with the Berry curvature as well as the quantum metric (through the Christoffel symbols of the second kind).
Combined with the geometric property of the injection current in Eq.~\eqref{eq:inj-geometry}, it shows that $PT$-symmetric responses originate from the quantum metric only and $T$-symmetric responses are controlled by both the Berry curvature and the quantum metric.
When the diamagnetic term $w^{ac}_{mn}$ in Eq.~\eqref{eq:shift-matrix} vanishes, the relation between the shift conductivity and the quantum metric and Berry curvature can be made more direct from
\begin{align}
\label{eq:shift-geometry-linear}
\sigma^{c;ab}_{\rm shift}
&=-\frac{i}{\omega}
\frac{2\pi e^3}{\hbar^2}
\int_{\omega_{cv}=\omega}
\frac{d^{d-1}k}{(2\pi)^d}
\left[(\hat{n}\cdot\hat{a})Q_{bc}
-(\hat{n}\cdot\hat{b})Q^*_{ac}
\right]
\end{align}
for $\omega>0$ when the diamagnetic term vanishes.
This formula can be applied, e.g., to Dirac and Weyl Hamiltonians that are at most linear in momentum.
Note that the real and imaginary part of the conductivity in Eq.~\eqref{eq:shift-geometry-linear} has the form of the Berry curvature dipole~\cite{sodemann2015quantum} and the quantum metric dipole~\cite{gao2019nonreciprocal}, respectively.

\subsection{Generalization to Multibands}

Let us discuss generalizing our geometric interpretation to include three or more bands
(when bands are doubly degenerate due to $PT$ symmetry, this means that we consider six or more bands).
The shift and injection conductivity takes the form
$\sum_{n\in {\rm occ}}\sum_{m\in{\rm unocc}}\int_{\bf k}I^{c;ab}_{nm}\delta(\omega-\omega_{mn})$ for $\omega>0$.
Because the energy conservation imposed by the delta function chooses a particular set of an unoccupied band $m$ for an occupied band $n$, the interband-transitive optical response is, in general, not associated with a property of the occupied band alone.
On the other hand, for example, the quantum geometric tensor $Q_{ba}$ is defined by suming over all occupied $n$ and unoccupied $m$ indices of the matrix elements by $\sum_{n\in \rm occ}\sum_{m\in \rm unocc}r^{b}_{nm}r^{a}_{mn}$, and so it becomes a property of the ground state $\sum_{n\in \rm occ}\sum_{m\in \rm all}r^{b}_{nm}r^{a}_{mn}-\sum_{n\in \rm occ}\sum_{m\in \rm occ}r^{b}_{nm}r^{a}_{mn}$, depending only on the occupied states.

In our analysis, though, we focus on Dirac and Weyl points where the quantum geometric tensor diverges at the gap-closing points.
Thus, the geometric quantities of the occupied bands are dominated by the property of the two crossing bands $n=1$ and $m=2$, through a large value of $r^{b}_{12}r^{a}_{21}$ and their derivatives, and they manifest through the shift and injection currents for small $\omega$.
Similarly, when bands are Kramers degenerate due to $PT$ symmetry, the matrix elements involving the indices for the four crossing bands are dominant contributions.
It means that, at low frequencies, we have a good {\it geometric approximation} for the conductivity tensors by
\begin{align}
\label{eq:geometric-approximation}
\sigma^{c;ab}_{\rm inj}(\omega)
&\approx
-\tau \frac{2\pi e^3}{\hbar^2}
\sum_{n,m}
\int_{{\bf k}:\omega_{mn}=\omega}
f^{\rm FD}_{nm}(\hat{n}\cdot \hat{c})Q_{ba},\notag\\
\sigma^{c;ab}_{\rm shift}(\omega)
&\approx
-i\frac{2\pi e^3}{\hbar^2}
\int_{\bf k}
\sum_{n,m}
f^{\rm FD}_{nm}C_{bca}
\delta(\omega_{mn}-\omega),
\end{align}
where $Q_{ba}$ and $C_{bca}$ are quantum geometric tensor and quantum geometric connection, respectively, defined by
\begin{align}
Q_{ba}
&=
\sum_{n\in \rm occ}\sum_{m\in\rm unocc}r^b_{nm}r^a_{mn},\notag\\
C_{bca}
&=\frac{1}{2}Q_{bd}(g^{-1})^{de}\left(\d_cg_{ba}+\d_ag_{bc}-\d_bg_{ca}\right),
\end{align}
where $g_{ba}={\rm Re}\left[Q_{ba}\right]$ as above.
In general, the injection conductivity tensors Eq.~\eqref{eq:geometric-approximation} differs from the exact expression in Eq.~\eqref{eq:shift-inj} because the former has the information of all band indices rather than the specific bands $n$ and $m$ involved in the optical transition.
Moreover, additional differences come in the shift conductivity tensors due to the {\it virtual transition terms}: $C_{bca}=-i\sum_{n\in \rm occ}\sum_{m\in \rm unocc}R^{c,a}_{mn}r^b_{nm}r^a_{mn}$+virtual transitions (See Appendix~\ref{sec:multiband}).
When the optical excitation occurs very close to a gap-closing point, however, only the band indices near the gap closing significantly contribute to the geometric quantities, effectively selecting specific band indices.
Also, virtual transition terms are suppressed by a factor $(\omega/\Delta E)^2$~\cite{de2017quantized}, where $\Delta E$ is the characteristic energy difference between the crossing bands and the other bands.
Thus, Eq.~\eqref{eq:geometric-approximation} becomes a good approximation near the gap closing.
Let us note that, in two-band or $PT$-symmetric four-band models, Eq.~\eqref{eq:geometric-approximation} becomes exact and corresponds to expressions Eqs.~\eqref{eq:inj-geometry} and~\eqref{eq:shift-geometry} above.

On the other hand, insulators or ordinary metals do not have geometric singularities in general, and the geometric approximation Eq.~\eqref{eq:geometric-approximation} does not apply to them so well.
Nevertheless, recalling that the Berry curvature of each band (rather than that of the whole occupied bands) is well-defined, we can hope for a possibility of defining a well-defined geometric quantity associated with a pair of bands also.
Let us see whether it makes sense to give a geometric meaning to the matrix element $r^a_{nm}r^b_{mn}\equiv g_{ab;nm}-iF_{ab;nm}/2$ by focusing on the real part $g_{ab;nm}$ (we note that it is different from the nonabelian quantum metric~\cite{rezakhani2010intrinsic,ma2010abelian} of the occupied bands, defined by $(g_{ab})_{n_1n_2}=\sum_{m\in \rm unocc} (r^a_{n_1m}r^b_{mn_2}$).
Since $g_{ab;nm}$ is a positive-semidefinite symmetric rank-2 tensor. i.e., $g_{aa;nm}\ge 0$ for all $a$ for given $n$ and $m$, this quantity is meaningful as a metric tensor although its interpretation is not clear.
This point of view can help to understand the structure of the circular shift current.
One can see that the matrix element of the circular shift current $R^{c,a}_{nm}r^b_{nm}r^a_{mn}$ can be written as $\Gamma_{bca;nm}+\text{virtual transitions}$, where $\Gamma_{bca;nm}$ is the Christoffel symbol of the first kind defined from the metric $g_{ba;nm}$ (See Appendix~\ref{sec:multiband} for a derivation).
Therefore, one may still regard the shift current as originating from a Christoffel-symbol-like quantity, when the virtual transitions, terms involving virtual transitions, is negligible.

\section{Model calculations}
\label{sec:model}

Our theoretical analysis reveals the circular shift current as an interesting new component of the interband bulk photovoltaic response in magnetic systems.
Also, the full generality of our theory allows us to understand the shift and injection currents in Dirac and Weyl semimetals in arbitrary spatial dimensions in a unified way.
In this section, we investigate the shift and injection current responses of tilted Dirac and Weyl points more closely with explicit calculations of the conductivity tensors.
We first deal with massless Dirac and Weyl Hamiltonians up to linear order in momentum, which cover both type-I and type-II spectra in arbitrary spatial dimensions.
In addition to the symmetry and divergence properties investigated above, we show the phenomenon of {\it separation of responses}, meaning that nonvanishing $PT$-symmetric responses and the $T$-symmetric responses do not coexist in the same component.
It can occur generically in tilted Dirac and Weyl systems having $C_{2x}$ or $M_{x}T$ symmetry, where $x$ is the direction of tilting.
Next, we study a model of Dirac surface states of magnetic topological insulators, which includes $k^2$ and $k^3$ terms in the Hamiltonian.

\subsection{Tilted Weyl and Dirac semimetals: ${\bf k}$-linear order}

Let us first revisit the model of a tilted Weyl point~\cite{chan2017photocurrents,yang2017divergent} to understand its general  second-optical response in more detail.
The Hamiltonian has the form
\begin{align}
\label{eq:tilted-Weyl}
H_{\rm Weyl}
=-\mu-\hbar v'k_x
+\hbar v(k_x\sigma_x+k_y\sigma_y+k_z\sigma_z).
\end{align}
The Weyl point described by this Hamiltonian is called type-I when $|v'/v|<1$ and type-II when $|v'/v|>1$~\cite{soluyanov2015type}.
The spectrum for the two cases are shown in Fig.~\ref{fig:tilted}(b,c).
When light with frequency $\omega>0$ is illuminated, optical excitation occur when two bands have energy difference $\hbar\omega_{cv}=2\hbar vk=\hbar \omega$ due to the energy conservation and only the lower band is occupied, i.e.,
$\hbar \omega_v=-\mu+v'k \cos\theta-vk<0$ and $\hbar\omega_c=-\mu+v'k \cos\theta+vk>0$ (Fig.~\ref{fig:tilted}), where $k=|{\bf k}|$ and $k_x=k\cos\theta$.
When $v'\ne 0$, the range of $\theta$ does not cover the whole sphere and is confined to a subspace $\theta_-<\theta<\theta_+$ in general [Fig.~\ref{fig:tilted}], where $\theta_{\pm}$ are functions of dimensionless parameters $2\mu/\hbar\omega$ and $v'/v$.
The minimal angle $\theta_-$ is always zero for a type-I Weyl point, but it is typically nonzero for a type-II Weyl point [Fig.~\ref{fig:tilted}(d,e)].
This asymmetric excitation leads to a nonzero optical conductivity given by
\begin{align}
\label{eq:tilted-Dirac-transitive}
\sigma^{c;ab}_{\rm shift}
&=
\frac{e^3}{2\pi \hbar^2}
\frac{1}{\omega}
\left(\frac{v}{\omega}\right)^{3-d}
\left[{\cal F}^{c;ab}_{\rm shift}(\theta_+)-{\cal F}^{c;ab}_{\rm shift}(\theta_-)\right]\notag\\
\sigma^{c;ab}_{\rm inj}
&=
\frac{e^3}{2\pi \hbar^2}
\tau
\left(\frac{v}{\omega}\right)^{3-d}
\left[{\cal F}^{c;ab}_{\rm inj}(\theta_+)-{\cal F}^{c;ab}_{\rm inj}(\theta_-)\right].
\end{align}
The form of ${\cal F}^{c;ab}(\theta)$ for nonvanishing components are summarized in Table.~\ref{tab:tilted}.
Because of the SO(2) rotational symmetry around the $x$ axis, there are four independent components 
\begin{align}
\sigma^{x;xx}_{\rm L}, 
\sigma^{x;yy}_{\rm L}=\sigma^{x;zz}_{\rm L}, 
\sigma^{y;xy}_{\rm L}=\sigma^{z;xz}_{\rm L},
 \sigma^{y;zx}_{\rm L}=-\sigma^{z;yx}_{\rm L}
\end{align}
for the real component, and three independent components
\begin{align}
\sigma^{x;yz}_{\rm C},
\sigma^{y;xy}_{\rm C}=\sigma^{z;xz}_{\rm C},
\sigma^{y;zx}_{\rm L}=-\sigma^{z;yz}_{\rm C}
\end{align}
for the imaginary component of the conductivity.
${\cal F}^{c;ab}(\theta)$ follows the same symmetry.
Remarkably, $PT$-symmetric responses (linear injection and circular shift) and $T$-symmetric responses (linear shift and circular injection) do not coexist in the same component.
To understand this, let us note that $PT=C_{2y}M_yT=C_{2z}M_zT$.
Since our model has $C_{2y}T$ and $C_{2z}T$ symmetries, nonvanishing $PT$-symmetric responses appear in the components that are invariant under $M_y$ and $M_z$, whereas $T$-symmetric responses appear in the components that reverses sign under $M_y$ and $M_z$.

\begin{figure*}[!t]
\includegraphics[width=\textwidth]{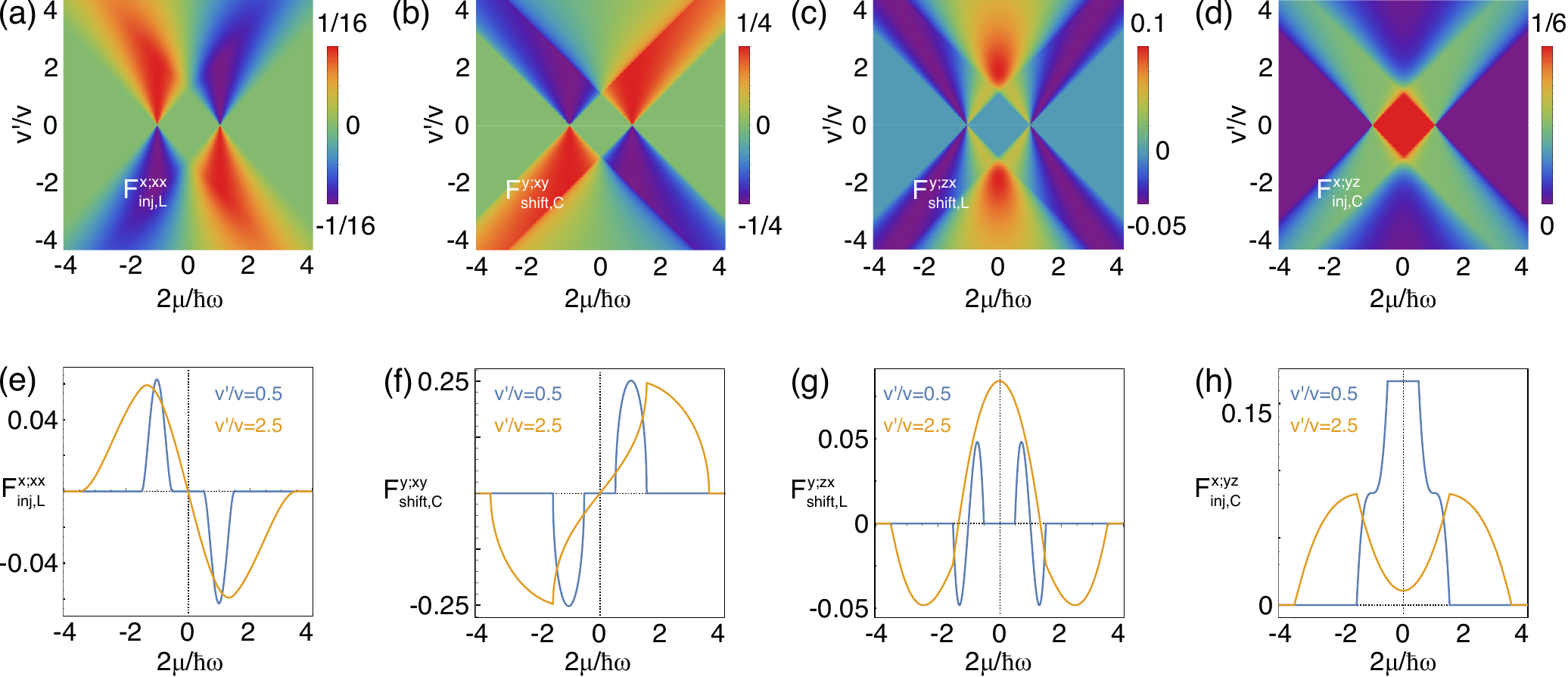}
\caption{${\cal F}^{c;ab}$ of a tilted Weyl point in three dimensions.
(a,e) $x;xx$ component of the linear injection.
(b,f) $y;xy$ component of the circular shift.
(c,g) $y;zx$ component of the linear shift.
(d,h) $x;yz$ component of the circular injection.
Since we take $\omega>0$, the sign of $2\mu/\hbar\omega$ here should be understood as the sign of the chemical potential $\mu$.
}
\label{fig:Dirac-numerical}
\end{figure*}

Figure~\ref{fig:Dirac-numerical} shows some representative components calculated from quantum geometric quantities by Eqs.~\eqref{eq:inj-geometry},~\eqref{eq:QGT}, and~\eqref{eq:shift-geometry-linear}.
There are some features that need to be discussed.
First, as we explained in Sec.~\ref{sec:symmetry} only the circular injection current is nonvanishing at the neutral filling $\mu=0$ in the type-I case where $|v'/v|<1$.
It is because, in this case, the Fermi surface is a point, so anisotropic excitation cannot occur.
The circular injection response is quantized because of the quantized Berry curavature from a Weyl point, as found in Ref.~\cite{de2017quantized} [See Eq. ~\eqref{eq:inj-geometry}].
Other responses are significant near $\hbar\omega=2\mu$, where the anisotropic excitation occurs.

However, there are significant differences between the $PT$-symmetric responses and the linear shift response.
$PT$-symmetric responses have peaks at $\hbar\omega=|2\mu|$, i.e., when the excitation occurs on a full hemisphere, while $T$-symmetric responses vanish at $\hbar\omega=|2\mu|$ and changes sign there~\cite{yang2017divergent}.
For the linear injection current, the peak at $\hbar\omega=|2\mu|$ is natural because the transition of the electron velocity during the excitation from the valence band $v$ to the conduction band $c$, $\Delta^{x}_{cv}=v^x_c-v^x_v$, is all positive or all negative on the hemisphere.
There is no simple analogous way to understand the trend of shift current response based on the shift vector, but the vanishing of the linear shift current response at $\hbar\omega=|2\mu|$ can be attributed to the $T$-symmetric nature.
Since $T$ symmetry requires that the current generated from one hemisphere equal to the current generated from the other hemisphere, both hemisphere should generate zero currents because linear shift currents are not generated when excitations occur on the full sphere.
There are also interesting differences between the $PT$-symmetric responses and $T$-symmetric responses at $\mu=0$ in the type-II case where $|v'/v|>1$.
As explained in Sec.~\ref{sec:symmetry}, emergent $CPT$ symmetry at $\mu=0$ requires that the former has a vanishing response while the latter can have a nontrivial response.
In other words, $PT$-symmetric responses do not distinguish type-I and type-II cones at $\mu=0$ whereas $T$-symmetric responses distinguish them.

As we understand a single Weyl point, it is straightforward to extend our knowledge to Dirac points in two and three dimensions.
In both 2D and 3D, the protection of a Dirac point requires $PT$ symmetry.
In 2D, the Dirac Hamiltonian has the form
$H_{\rm 2D}=-\mu-\hbar v'k_x+\hbar v(k_x\tau_x+k_y\tau_y)$,
where $\tau_{i=x,y,z}$ are Pauli matrices for the orbital degrees of freedom, and spinless $PT=\sigma_zK$ symmetry forbids the mass term $m\sigma_z$.
In 3D, the Dirac Hamiltonian has the form
$H_{\rm 3D}=-\mu-\hbar v'k_x+\hbar v(k_x\tau_x+k_y\tau_y\sigma_x+k_z\tau_z)$,
and the twofold degeneracy (Kramers degeneracy) of bands at every momentum require $PT=i\sigma_yK$ symmetry.
Here, $\tau_{i=x,y,z}$ and $\sigma_{i=x,y,z}$ are Pauli matrices for the orbital and spin degrees of freedom, respectively.
In 3D, even in the presence of $PT$ symmetry, two mass terms are allowed, which are $m_1\tau_y\sigma_y$ and $m_2\tau_y\sigma_z$.
We need threefold or fourfold rotational symmetry to protect the Dirac point in 3D by disallowing mass terms.
Our massless Dirac model have continuous $\theta$ rotational symmetry around the $x$ axis under $C_{\theta}=e^{i\theta(\sigma_x+\tau_x\sigma_x/2)}$, satisfying $C_{\theta}H({\bf k})C_{\theta}^{-1}=H(C_{\theta}{\bf k})$.
Keeping either threefold $C_{3x}$ or fourfold $C_{4x}$ rotational symmetry in crystals preserves the gap closing~\cite{wang2012dirac,wang2013three,yang2014classification}, which we assume here.
Because gapless Dirac points have $PT$ symmetry, they can only have linear injection or circular shift current responses.
These responses have the same pattern shown for a Weyl point.

Here we emphasize again that while multiple responses coexist in magnetic Weyl and Dirac semimetals, each response occurs through different components of the conductivity.
This helps measure each response separately.
In particular, it facilitates the measurement of the circular shift current in magnetic Dirac and Weyl semimetals.
Table.~\ref{tab:tilted} and Eq.~\eqref{eq:currents} shows that the current generated along the $y$ direction under the illumination of circularly polarized light propagating in the $z$ direction is only the circular shift currents.
The photocurrent along the $y$ direction should thus be identified with the circular shift current.

\subsection{Dirac surface state}

As an application to a more realistic model with ${\bf k}$-nonlinear terms, we study the single Dirac surface state of magnetic topological insulators.
Let us begin with the following effective Hamiltonian studied in Refs.~\cite{semenov2012tunable,ogawa2016zero}.
\begin{align}
H
&=
-\mu+\frac{\hbar^2k^2}{2m}
+
\hbar v\left(k_x\sigma_y-k_y\sigma_x\right)+\Delta\sigma_y.
\end{align}
Here, $\Delta\ne0$ is due to spin ordering along the $y$ direction, and it breaks $M_x=i\sigma_x$, rotation $C_{2z}=-i\sigma_z$, and time reversal $T=i\sigma_yK$ symmetries.
Since this in-plane ordering preserves $C_{2z}T$ symmetry, it does not open the gap, and it just shifts the location of Dirac points by $-\Delta/\hbar v$ from the time-reversal-invariant momentum.
The shifting tilts the Dirac cone because of the quadratic term: if we write $(k_x,k_y)=(-\Delta/\hbar v+q_x,q_y)$, the Hamiltonian has the form $H=-\mu-(\hbar\Delta/mv)q_x+\hbar v\left(q_x\sigma_y-q_y\sigma_x\right)$ up to linear order in $q$, which is studied above.
Assuming $C_{3z}$ symmetry of the nonmagnetic state, we add a hexagonal warping term in order to account for the crystalline symmetry of the real system.
\begin{align}
h_{\rm warp}
&=\lambda(k_x^3-3k_xk_y^2)\sigma_z.
\end{align}
This term breaks $M_y=i\sigma_y$ symmetry and $C_{2z}T$ symmetry that are preserved by the spin ordering, so it opens small band gap (about $0.8$ meV for parameters given below).
Since $T$ and $C_{2z}T$ symmetries are both broken, all four types of shift and injection currents can be generated in this system.
However, the residual $M_xT$ symmetry imposes that the separation of responses remains exact: nonvanishing components of the conductivity are $\sigma^{y;xx}_{\rm shift,L}$, $\sigma^{y;yy}_{\rm shift,L}$, $\sigma^{x;xy}_{\rm shift,L}$ for linear shift current, $\sigma^{x;xx}_{\rm inj,L}$, $\sigma^{x;yy}_{\rm inj,L}$, $\sigma^{y;xy}_{\rm inj,L}$ for linear injection current, $\sigma^{y;xy}_{\rm shift,C}$ for circular shift current, and $\sigma^{x;xy}_{\rm inj,C}$ for circular injection current.

\begin{figure}[t!]
\includegraphics[width=8.5cm]{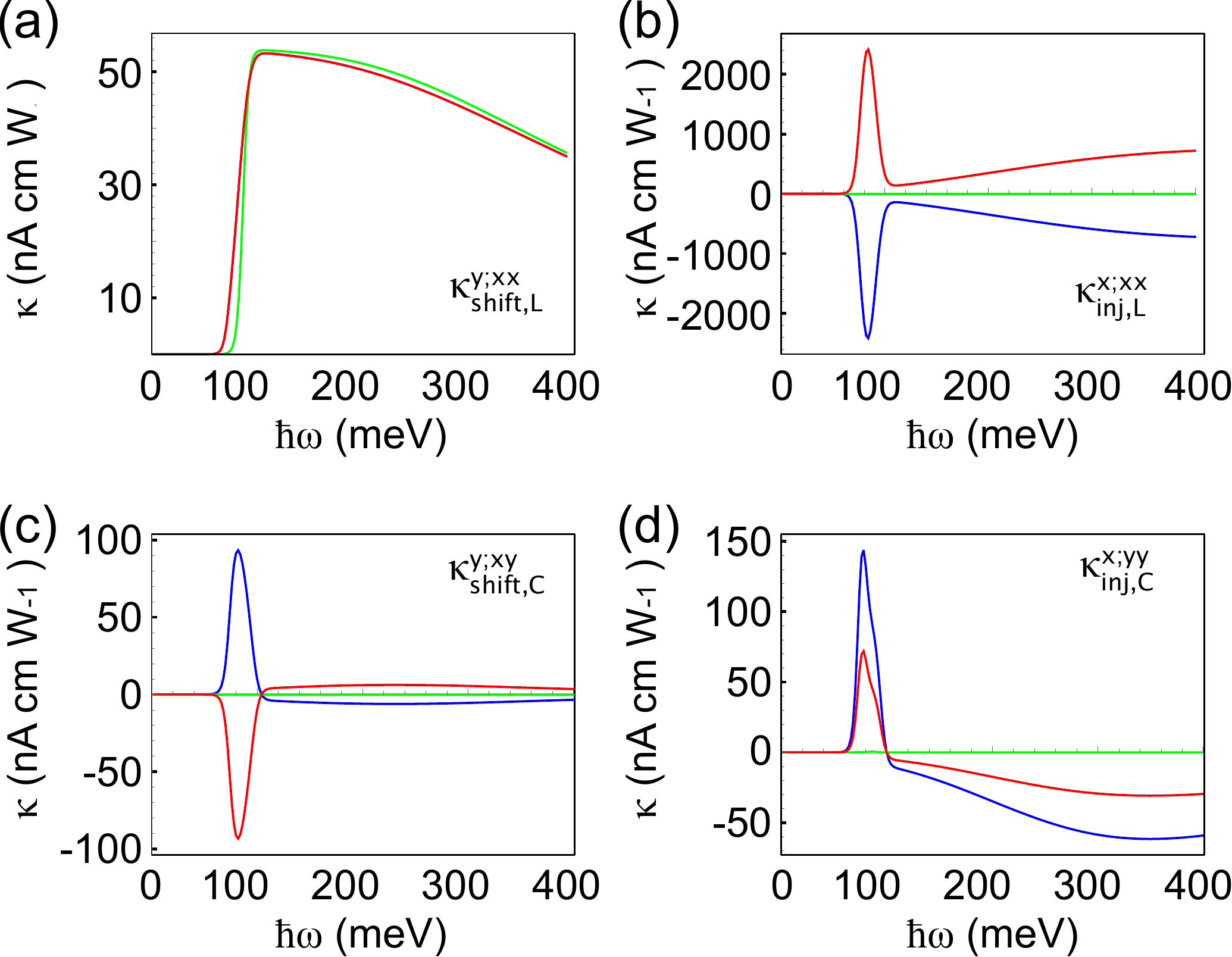}
\caption{
Photoresponsivity of the magnetic Dirac surface state.
$\mu=50$ meV,
$m=0.13 m_e$,
$\hbar v=2.5\;{\rm eV\AA^{-1}}$,
$\lambda=250\;{\rm eV\AA^{-3}}$,
and $\hbar \tau^{-1}=1$ meV.
Blue, green, and red curves correspond to $\Delta=30$ meV, $0$ meV, and $-30$ meV, respectively.
$\kappa^{c;ab}=1\;{\rm nA cm W^{-1}}$ is equivalent to $\sigma^{c;ab}=1.33\;{\rm pAcm^{-1}V^{-2}}$.
}
\label{fig:numerical}
\end{figure}

For a numerical calculation, we take $\mu=50$ meV, $m=0.13m_e$ where $m_e$ is the free electron mass, $\hbar v=2.5\; {\rm eV\AA^{-1}}$, $\lambda=250\;{\rm eV\AA^{-3}}$, and $\hbar \tau^{-1}=1$ meV and use Eqs.~\eqref{eq:inj-geometry} and~\eqref{eq:shift-geometry}.
Figure~\ref{fig:numerical} show the calculated photoresponsivity $\kappa^{c;ab}=2\sigma^{c;ab}/\epsilon_0c$, which has the dimension of the photocurrent density per unit intensity of light~\cite{cook2017design}.
The peak value (occurring at $\hbar\omega\sim 2\mu=100$ meV) of the linear injection part is the strongest, and the others are smaller by two orders of magnitude.
However, since the circular shift current grows as $\omega^{-2}$ while the linear injection current grows as $\omega^{-1}$, the circular shift current can be comparable to or larger than the linear injection current when the peak is located below $10$ meV.
On the other hand, the small-frequency divergence of the linear shift current is weaker because it is due to $\lambda\ne 0$~\cite{kim2017shift} and is not from the tilting, and thus the peak value scales like $O(\omega^{0})$ as $\mu$ is lowered.
Therefore, the $y$ component photocurrent generated by a circularly polarized light, $j^y=(\kappa^{yxx}_{\rm shift,L}+\kappa^{yyy}_{\rm shift,L}-2\kappa^{y;xy}_{\rm shift,C})I$, is dominated by the circular shift ($\kappa^{y;xy}_{\rm shift,C}$) current when $\hbar\omega\sim2\mu<100$ meV.
The magnitude of the linear shift current and the circular shift current can be compared in experiments since the circular parts can be measured from the current difference between the left-circularly polarized light and the right-circularly polarized light.

In experiments, the value of the observed photocurrents can be smaller than the value predicted here.
For example, the photocurrents observed in Ref.~\cite{ogawa2016zero} shows photoresponsivity of about $5\;{\rm nAcm^{-1}W^{-1}}$ at $\omega\sim 250$ meV, which is two orders of magnitude smaller than the calculated value here.
While several factors can contribute to this reduction, one is from the cancellation between the top and bottom surfaces.
This cancellation can be reduced by increasing the thickness of the sample because light attenuates more while propagating within the bulk such that the photocurrent is generated significantly on only one surface that is directly illuminated.

\section{First-principles calculations for real topological semimetals}
\label{sec:first-principles}

For further demonstrations of our theory beyond simple two-band models,
we perform first-principles 
calculations on the shift and injection photocurrent conductivities as well as geometric
quantities of antiferromagnetic MnGeO$_3$ and ferromagnetic PrGeAl, respectively,
as representatives of real magnetic Dirac and Weyl semimetals.
We notice that such calculations on the bulk photovoltaic effects in real magnetic topological semimetal have not been reported yet
despite the fact that topological semimetals are expected to be efficient infrared and terahertz photodetectors~\cite{liu2020semimetals}.

\subsection{Antiferromagnetic Dirac semimetal MnGeO$_3$}
MnGeO$_3$ forms a centrosymmetric rhombohedral structure [see Fig.~\ref{fig:structure}(a)] with space group
$R\overline{3}$~\cite{tsuzuki1974neutron}, and consequently, it would not exhibit any bulk photovoltaic effects. 
Interestingly, it becomes antiferromagnetic below 38 K~\cite{tsuzuki1974neutron}
and the AF structure (magnetic space group $-3'$) [see Fig.~\ref{fig:structure}(a)]
breaks both $T$ and $P$ symmetries while preserving the combined $PT$ symmetry~\cite{xu2020high},
thus leading to AF-induced bulk photovoltaic effects with linear injection and circular shift currents (see Table I).
Furthermore, it was recently predicted to be a Dirac semimetal with $PT$ symmetry-protected Dirac points (DPs)~\cite{xu2020high}. 

\begin{figure}[t] \centering
\includegraphics[width=8.5cm]{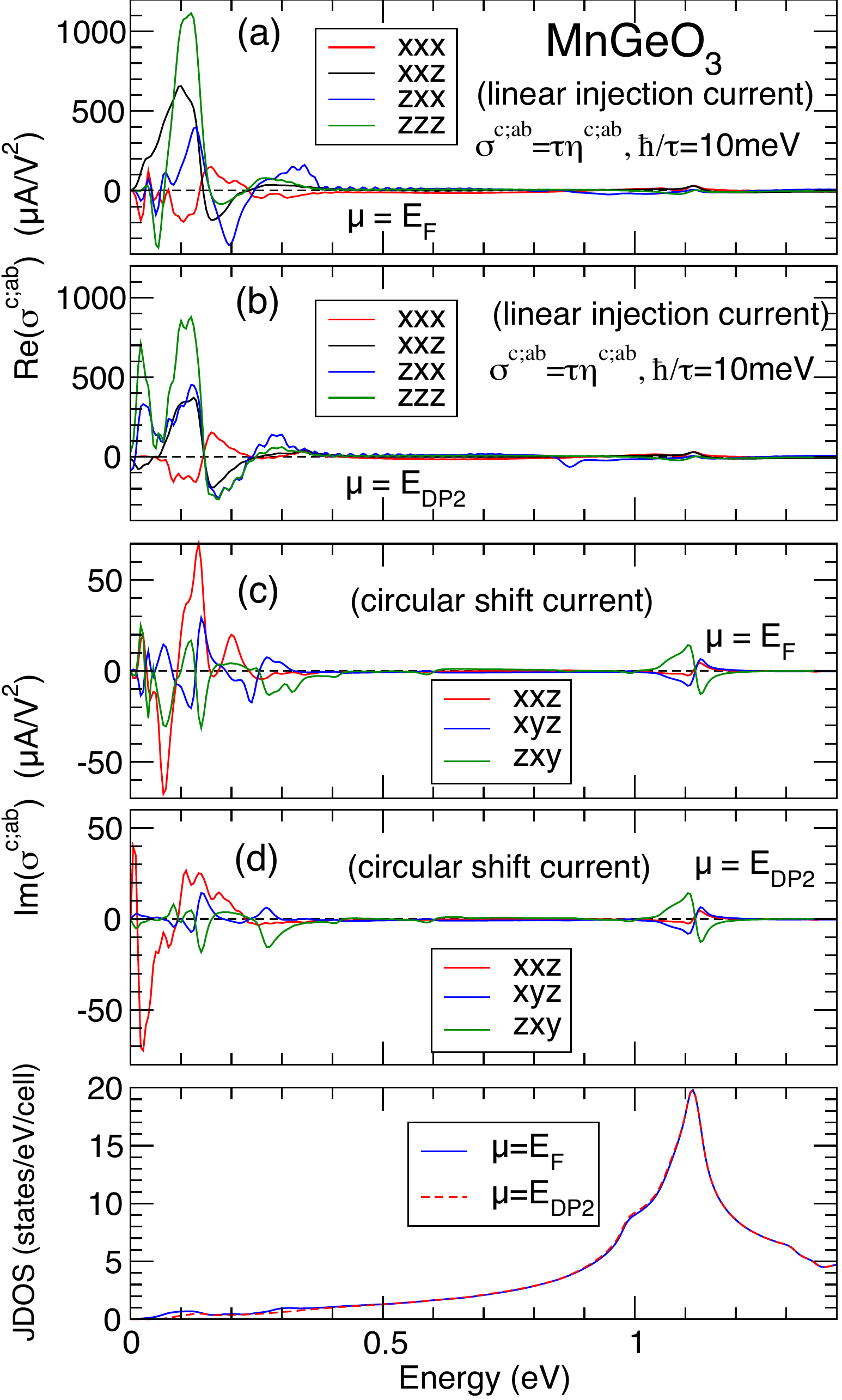}
\caption{Conductivity tensors and joint density of states (JDOS) of antiferromagnetic MnGeO$_3$. (a, b) Linear injection. (c, d) Circular shift.
In (b) and (d), the conductivities are calculated with
the Fermi level set to the DP$_2$ Dirac point energy (see Fig.~\ref{fig:geom_MnGeO3}).
(e) JDOS as a function of photon energy.
}
\label{fig:sigma_MnGeO3}
\end{figure}

In AF MnGeO$_3$, because of its $PT$ symmetry, there are only nonvanishing
circular shift photocurrent and linear injection photocurrent, as mentioned before (see Table I).
Furthermore, the $-3'$ magnetic space group admits only three nonvanishing independent matrix elements 
(i.e., $xxz=-xzx=yzy$, $xyz=-xzy=-yxz$ and $zxy=-zyx$) of the circular shift conductivity tensor and
six nonvanishing independent matrix elements (i.e., $xxx=-xyy=-yxy$, $xyz=-yxz$, $xxz=yyz$, $xxy=yxx=-yyy$, $zxx=zyy$ and $zzz$) 
of the linear injection conductivity tensor~\cite{gallego2019automatic}.
Hereafter we use the shorthand notation $cab$ for $\sigma^{c;ab}$.
We display these nonvanishing conductivity elements in the low photon energy range
in Fig.~\ref{fig:sigma_MnGeO3}. For simplicity, we plot only the four pronounced $xxx$, $xxz$, $zxx$ and $zzz$
elements of the linear injection conductivity tensor in Fig.~\ref{fig:sigma_MnGeO3}.
$\hbar\tau^{-1}$ = 10 meV is assumed.
We notice that the magnitudes of the linear injection conductivity elements ($\sigma^{c;ab}$)
are gigantic in the photon energies below 0.25 eV (Fig.~\ref{fig:sigma_MnGeO3}). 
The magnitudes are order of $\tau e^3/(2\pi\hbar^2)=500$ $\mu$A/V$^{2}$, which are one order of magnitude larger than those in architypical polar semiconductors CdS and CdSe~\cite{nastos2010optical}.
Circular shift photocurrents (Fig.~\ref{fig:sigma_MnGeO3})
are also 10 times larger than linear shift currents in semiconductors CdS and CdSe~\cite{nastos2010optical}. 
This is remarkable because it demonstrates that the AF magnetism-induced linear injection
and circular shift photocurrents, respectively, can be as large as circular injection
and linear shift currents in nonmagnetic noncentrosymmetric materials.
Furthermore, this means that the photocurrents in AF semimetals can be controlled via
manipulating the magnetism with, e.g., an applied magnetic field~\cite{zhang2019switchable}.

\begin{figure*}[t] \centering
\includegraphics[width=14.5cm]{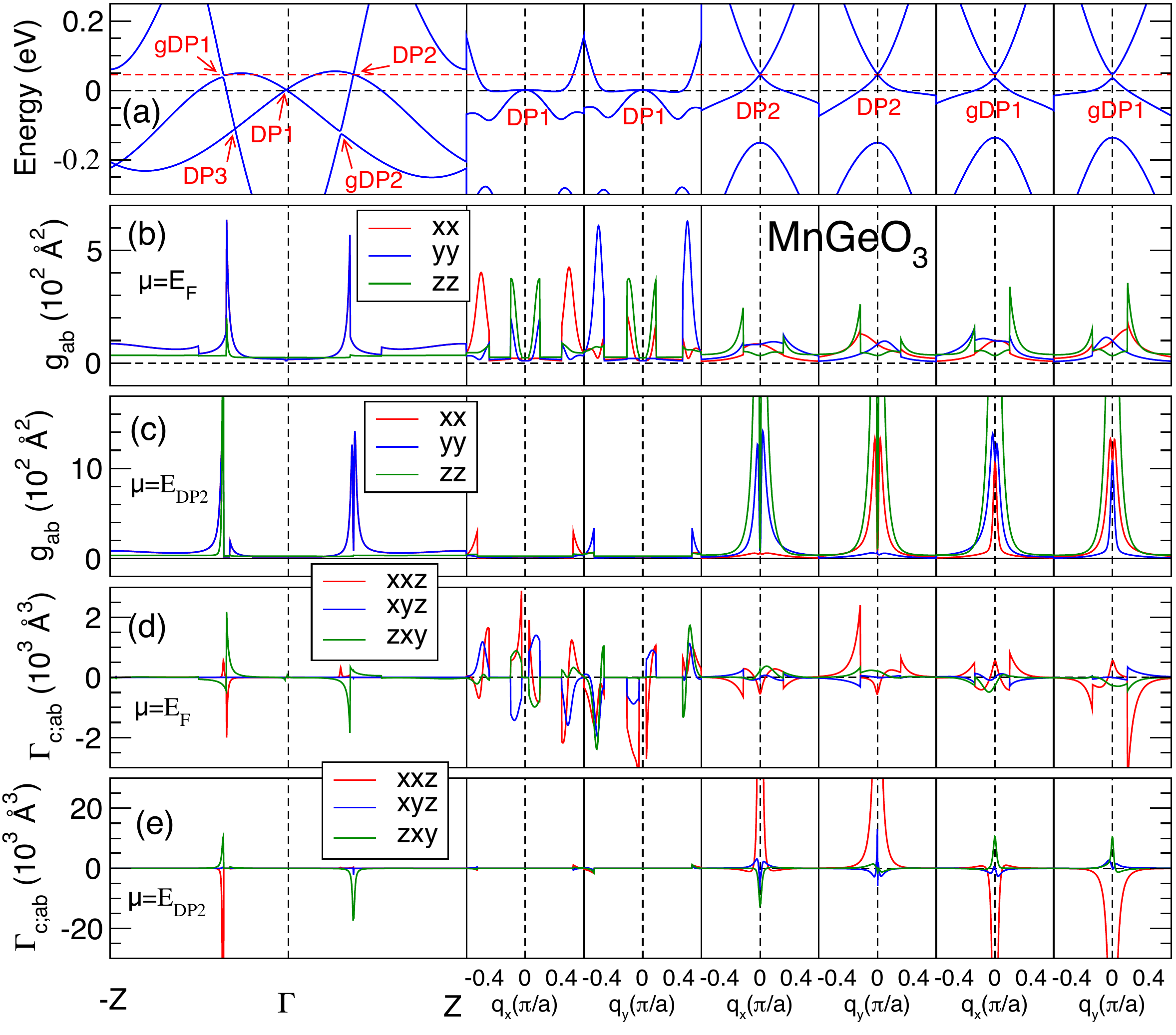}
\caption{Enhancement of geometric quantities near Dirac points.
(a) Energy bands along the $k_z$ axis from $-\pi/c$ to $\pi/c$ ($Z=\pi/c$) and also 
along the $k_x$ and $k_y$ directions through the DP1 and DP2 Dirac points as well as the 
gDP1 gapped Dirac point. (b) and (c) Quantum metric ($g_{ab}$) along the same symmetry lines as in (a),
calculated for the Fermi energy ($E_F$) and the DP2 energy ($E_{DP2}$), respectively. 
(d) and (e) The same as in (b) and (c), respectively, but for Christoffel symbol of the first kind 
($\Gamma_{c;ab}$). Here ${\bf q}={\bf k}-{\bf k}_{DP}$ denotes the momentum displacement from a DP (qDP) point.
}
\label{fig:geom_MnGeO3}
\end{figure*}

AF MnGeO$_3$ hosts at least three DPs near
the Fermi level $E_F$ along the $k_z$ axis, as shown in Fig.~\ref{fig:geom_MnGeO3}.
In particular, there is a DP just above $E_F$ (at 1.7 meV) and being located
close to the $\Gamma$ point [at ${\bf k}_{DP1}=(0,0,-0.00364)2\pi/a$].
This could explain the large values of the calculated photocurrents, as shown in Figs. \ref{fig:sigma_MnGeO3}(a)
and \ref{fig:sigma_MnGeO3}(c). To further examine the important contributions 
of the DPs to the photocurrents, we also calculate the conductivity spectra 
with the Fermi level set to the DP2 Dirac point energy (i.e., $E = 46.5$ meV). 
DP2 is located at ${\bf k}_{DP1}=(0,0,0.11112)2\pi/a$
above the $k_z = 0$ plane in the $k$-space [see Fig.~\ref{fig:geom_MnGeO3}(a)].
We notice that both the shapes and magnitudes of all the conductivity spectra
except conductivity element $\sigma^{xxz}$, roughly remain the same. 
For example, the gigantic peak of about 1000 $\mu$A/V$^2$ at $\sim$100 meV 
in the Re($\sigma^{zzz}$) linear injection current spectrum appears in both cases 
[see green curves in Figs.~\ref{fig:sigma_MnGeO3}(a) and~\ref{fig:sigma_MnGeO3}(b)]. 
Nonetheless, its sharp negative peak at 50 meV disappears in the case where the chemical potential $\mu$ is tuned to $\mu=E_{DP2}$
and a sharp positive peak of larger than 700 $\mu$A/V$^2$ occurs at 20 meV instead. 
Interestingly, there is a sharp positive peak at photon energy of 5 meV in
the Im($\sigma^{xxz}$) circular shift current calculated by setting $\mu=E_{DP2}$.
This is due to the $\omega^{-1}$ behavior of the shift conductivity near the Dirac point.
Comparing red curves in Figs.1 (c) and (d) for Im($\sigma^{xxz}$), one can also see that the large negative peak moves from photon energy 65 meV to 25 meV [see Fig.~\ref{fig:sigma_MnGeO3}(d)].

While an enhanced joint density of states (JDOS), $\rho(\omega)=\int_{\bf k}\sum_{m,n}f_{nm}\delta(\omega_{mn}-\omega)$, is a possible origin of large conductivity tensors in insulators~\cite{cook2017design}, it cannot explain the peaks shown here.
Figure~\ref{fig:sigma_MnGeO3}(e) shows that JDOS is suppressed at low-frequencies rather than being enhanced.
As we show above, linear injection and circular shift currents are, respectively, related to
geometric quantities quantum metric ($g_{ab}$) and Christoffel symbols of the first kind ($\Gamma_{c;ab}$) at low-energies through Eq.~\eqref{eq:geometric-approximation}.
Thus, the large enhancement of conductivity tensors has geometric origin.
To demonstrate this,
we display $g_{ab}$ and $\Gamma_{c;ab}$ at $\mu=E_F$ and also at $\mu=E_{DP2}$ 
along the symmetry lines in Fig.~\ref{fig:geom_MnGeO3}. 
For most of the DPs, a DP is associated with a gapped DP (gDP) located approximately at the inverted
position in the $k$-space. For example, the associated gPD for DP2 is gDP1 at ${\bf k}_{DP1}=(0,0,-0.11051)2\pi/a$,
and the DP2 energy level falls within the gDP1 band gap [see Fig.~\ref{fig:geom_MnGeO3}(a)]. 
On the other hand, there is no gDP associated with DP1. 
Figures \ref{fig:geom_MnGeO3}(b) and \ref{fig:geom_MnGeO3}(d) clearly show that for $\mu=E_F$, $g_{ab}$ and $\Gamma_{c;ab}$ have sharp peaks near DP1 along the $k_x$ and $k_y$ directions.
$g_{xx}$ and $g_{yy}$ also peak sharply at the positions of the DP2 and gDP1 along the $k_z$ axis
even though $\mu=E_F\neq E_{DP2}$.
This indicates that the gigantic linear injection and circular shift currents
stem, respectively, from the large values of the quantum metric and Christoffel symbol 
in the vicinity of the DP1 Dirac point. Furthermore, $g_{zz}$ has prominent peaks in the vicinity
of (but slightly away from) the DP2 and gDP1 points along the $k_x$ and $k_y$ directions,
which are mainly caused by the interband transitions from the lower (occupied) Dirac cone to
higher (empty) Dirac cone with transition energies of $\sim$0.1 eV [see Fig.~\ref{fig:geom_MnGeO3}(b)].  
These $g_{zz}$ peaks thus give rise to the gigantic peak in linear injection current Re($\sigma^{zzz}$)
at around 0.1 eV [see green curves in Fig.~\ref{fig:sigma_MnGeO3}(a)].

For $\mu=E_{DP2}$, as expected, 
$g_{ab}$ and $\Gamma_{c;ab}$ exhibit sharp peaks close to both the DP2 and gDP1 points 
[see Figs. \ref{fig:geom_MnGeO3}(c) and \ref{fig:geom_MnGeO3}(e)].
In particular, $g_{zz}$ has huge positive peaks near the DP2 and gDP1 points along all three cartesian coordinate
directions, thus resulting in the prominent peak at 0.1 eV in linear injection conductivity Re($\sigma^{zzz}$)
[Fig.~\ref{fig:sigma_MnGeO3}(b)], as in the $\mu = E_{F}$ case [Fig.~\ref{fig:sigma_MnGeO3}(a)].
Nevertheless, Fig.~\ref{fig:geom_MnGeO3}(e) shows that the $\Gamma_{c;ab}$ peaks near the DP2 and gDP1 have the opposite signs. 
This could explain why the circular shift conductivity elements
Im($\sigma^{xyz}$) and Im($\sigma^{zxy}$) become smaller when the Fermi level is raised from $E_F$ to $E_{DP2}$
[Figs. \ref{fig:sigma_MnGeO3}(c) and \ref{fig:sigma_MnGeO3}(d)] because the contributions from
the DP2 and gDP1 cancel each other to some extent. In contrast, the linear injection current elements
remain almost unchanged mainly because the $g_{ab}$ peaks have the same signs. 
Nonetheless, there is a huge peak in the $\Gamma_{xxz}$ at gDP1, which is absent at DP2,
because now the Fermi level falls within the gDP1 gap [see Fig.~\ref{fig:geom_MnGeO3}(e)]. 
This results in the large low energy peaks in the Im($\sigma^{xxz}$) conductivity element [see red curves in Fig.~\ref{fig:sigma_MnGeO3}(d)].

\subsection{Ferromagnetic Weyl semimetal PrGeAl}
PrGeAl forms a body-centered tetragonal structure (see Fig.~\ref{fig:structure}b)
with noncentrosymmetric space group $I4_1md$ and point group $4mm$.~\cite{gladyshevskii2000crystal}
It becomes ferromagnetic at $T_C = 16$ K~\cite{sanchez2020observation}  and the FM structure (Fig.~\ref{fig:structure}b) has
no $T$ symmetry nor $PT$ symmetry. 
Therefore, all four types of photocurrents may emerge in FM PrGeAl (Table I).
Interestingly, it was recently predicted to be a rare ferromagnetic
noncentrosymmetric Weyl semimetal~\cite{chang2018magnetic}. Furthermore, the Weyl nodes and surface Fermi arcs in PrGeAl
were observed in very recent photoemission spectroscopy experiments~\cite{sanchez2020observation}.
Thus, FM PrGeAl provides a valuable platform for studying all types of bulk photovoltaic effects
in Weyl semimetals.
\\

\begin{figure}[t] \centering
\includegraphics[width=8.5cm]{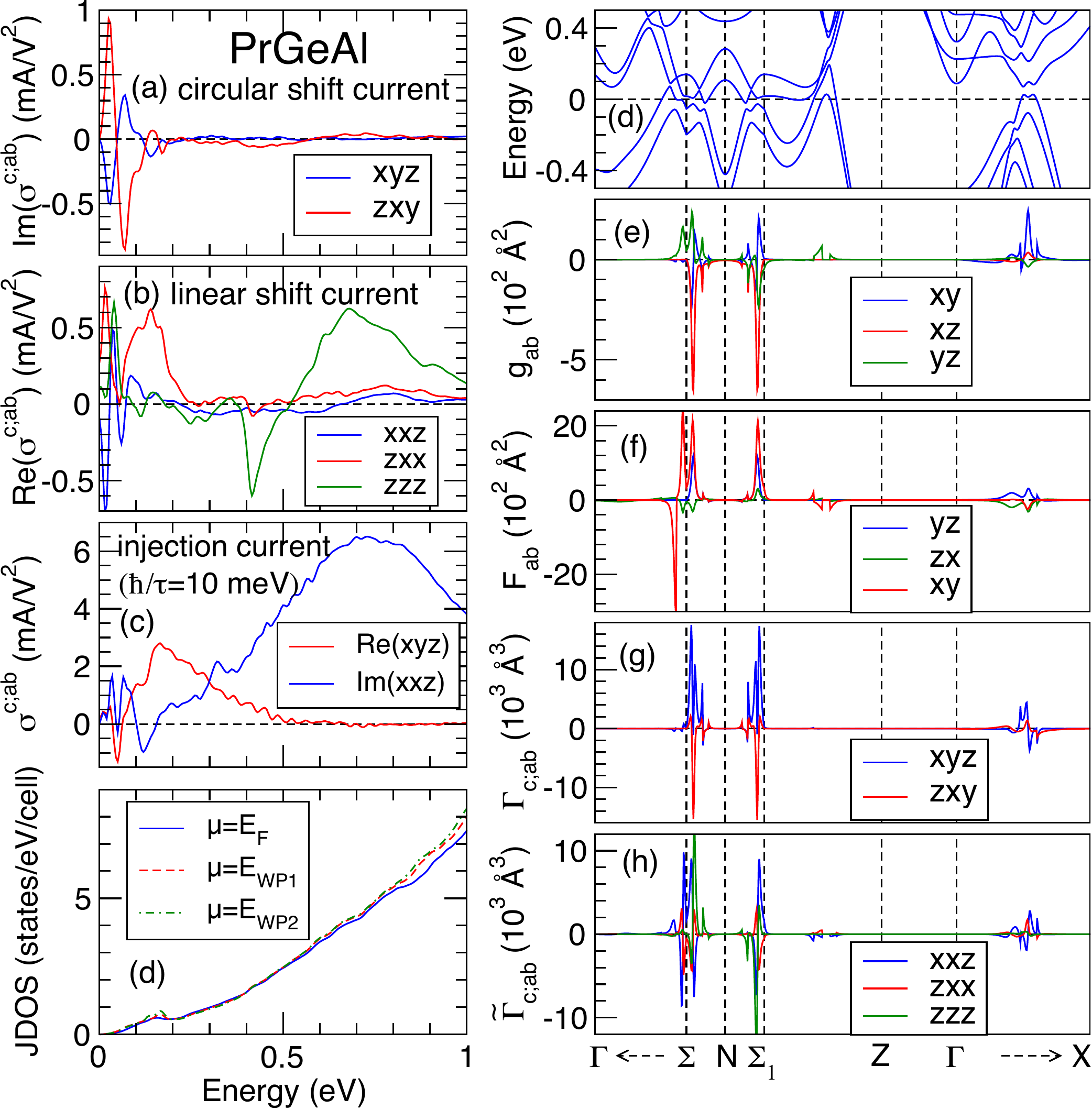}
\caption{
Conductivity tensors and geometric quantities in ferromagnetic PrGeAl.
(a, b) Shift and (c) Injection conductivity tensors.
(d) JDOS as a function of photon energy. The cases with $\mu=E_{\rm WP1}$ and $\mu=E_{\rm WP2}$ are also shown.
(e) Energy bands, (f) quantum metric $g_{ab}$, (g) Berry curvature $F_{ab}$, 
(h) Christoffel symbol $\Gamma_{c;ab}$, and (i) symplectic Christoffel symbol $\widetilde{\Gamma}_{c;ab}$ 
along the high symmetry lines in the Brillouin zone [see Fig.~\ref{fig:structure}(d) in Appendix~\ref{app:first-principles}].
}
\label{fig:sigma_PrGeAl}
\end{figure}

The crystalline point group of PrGeAl is $4mm$. Thus, there are three inequivalent nonvanishing matrix elements
(i.e., $xxz=yyz$, $zxx=zyy$ and $zzz$) of linear shift conductivity 
and one nonvanishing element ($xxz=yyz=-xzx=-yzy$) 
of circular injection conductivity in PrGeAl above $T_C = 16$ K.~\cite{gallego2019automatic}.
The magnetic point group of FM PrGeAl is $4m'm'$. Consequently, in addition, there are
two nonvanishing elements (i.e., $xyz=-xzy=-yxz$ and $zxy=-zyx$) of circular shift conductivity and one nonvanishing element
($xyz=-yxz$) of linear injection conductivity in PrGeAl below $T_C$.~\cite{gallego2019automatic}
Here, the presence of $M_xT$ and $M_yT$ symmetries forbids linear shift and injection conductivities 
to be simultaneously nonvanishing in the same component.
The same is true for the circular polarization.
The calculated conductivity spectra of these nonvanishing matrix elements are displayed in Fig.~\ref{fig:sigma_PrGeAl}.
We notice that all the shift current elements have large peaks below photon energy 0.25 eV
[see Figs. \ref{fig:sigma_PrGeAl}(a) and \ref{fig:sigma_PrGeAl}(b)].
This is due to the $\omega^{-1}$ enhancement of the shift conductivity tensors at low frequencies.
Indeed, the magnitudes of these peaks below 0.1 eV are comparable to 
those in architypical nonmagnetic Weyl semimetal TaAs~\cite{zhang2018photogalvanic},
which also has the $I4_1md$ space group.
Linear shift current element Re($\sigma^{zzz}$) is also large in the higher energy range between 0.4 eV and 1.0 eV. 
This peak is not due to magnetic order because it is observed in $T$-symmetric responses.
It is hardly related with the responses of Weyl points.
The peak is far from the Fermi level, and also the conductivity components showing the large peaks 
is not generated by linearly dispersing Weyl points [See Table.~\ref{tab:tilted} for example].
We do not aim to explain its origin here.
The calculated circular shift conductivity elements below 0.2 eV are comparable to that of the linear shift current 
(Fig. \ref{fig:sigma_PrGeAl}), and remarkably, are one order of magnitude larger than
those in AF Dirac semimetal MnGeO$_3$ [see Figs. \ref{fig:sigma_MnGeO3}(c) and \ref{fig:sigma_MnGeO3}(d)]. 
No first-principles calculation of circular shift current in other semimetals has been reported yet. 

Similar to the linear shift conductivity, circular injection conductivity Im($\sigma^{xxz}$) has
a gigantic broad peak between 0.5 eV and 1.0 eV, and the peak value is almost one order of magnitude
larger than that of linear injection conductivity in AF MnGeO$_3$ (see Fig.~\ref{fig:sigma_MnGeO3}).
It also has large peaks below 0.15 eV, although the magnitudes of these peaks are a few times smaller
than the gigantic peak in the higher energy region. Linear injection conductivity Re($\sigma^{xyz}$)
in FM PrGeAl is also large and is about ten times larger than that of AF MnGeO$_3$ in the very low energy region.
This one order of magnitude enhancement can be related to the presence of many Weyl points, 
as we analyze further below.

As in MnGeO$_3$, JDOS is suppressed at low energies [Fig.~\ref{fig:sigma_PrGeAl}(d)]
Thus, the origin of the large low-frequency photocurrents should be attributed to the geometric enhancement.
We have calculated all four geometric quantities along the symmetry lines in the Brillouin zone [Fig.~\ref{fig:structure}(d)], 
as displayed in Fig.~\ref{fig:sigma_PrGeAl}. We find that all four quantities have sharp peaks 
in the vicinity of the $\Sigma$ and $\Sigma_1$ points, where there are several anticrossing nodal points
with the Fermi level falling within their gaps [see Fig.~\ref{fig:sigma_PrGeAl}(e)].
Since these gaps near the anticrossing points are mostly within 0.1 eV, the large peaks in the geometric quantities
thus give rise to the sharp photoconductivity peaks around 0.1 eV and below. 
In particular, Christoffel symbols $\Gamma_{xyz}$ and $\Gamma_{zxy}$ have gigantic peaks near
the $\Sigma$ and $\Sigma_1$ points with the same signs, thus resulting in sharp peaks
in circular shift conductivities Im($\sigma^{xyz}$) and ($\sigma^{zxy}$), respectively, 
below photon energy 0.1 eV [Fig.~\ref{fig:sigma_PrGeAl}(a)]. 

FM PrGeAl has been reported to be a rare noncentrosymmetric FM Weyl semimetal
with at least 160 Weyl points (WPs) within $\pm$0.1 eV of the Fermi level~\cite{chang2018magnetic}.
Furthermore, there are both type I and II WPs among them.
Here we consider two WPs, one type I and the other type II, and study their influences on the photocurrents.
The type I WP (WP1) sits at ${\bf k}=(0.160,0.204,-0.003)2\pi/a$ in the BZ
and is located at 53.9 meV above $E_F$ ($E_{\rm WP1}$). The type II WP (WP2) is at ${\bf k}=(0.015,0.255,0.217)2\pi/a$ 
and its energy is 67.4 meV above $E_F$ ($E_{\rm WP2}$). They correspond, respectively, to Weyl points W$^1_3$ and W$^1_2$
reported in Ref. \cite{chang2018magnetic}. To study the influences of the WPs on the photocurrents,
we calculate all the nonvanishing conductivity elements and all the geometric quantities
with the Fermi level set to $E_{\rm WP1}$ and also to $E_{\rm WP2}$.  
\\

\begin{figure}[t] \centering
\includegraphics[width=8.5cm]{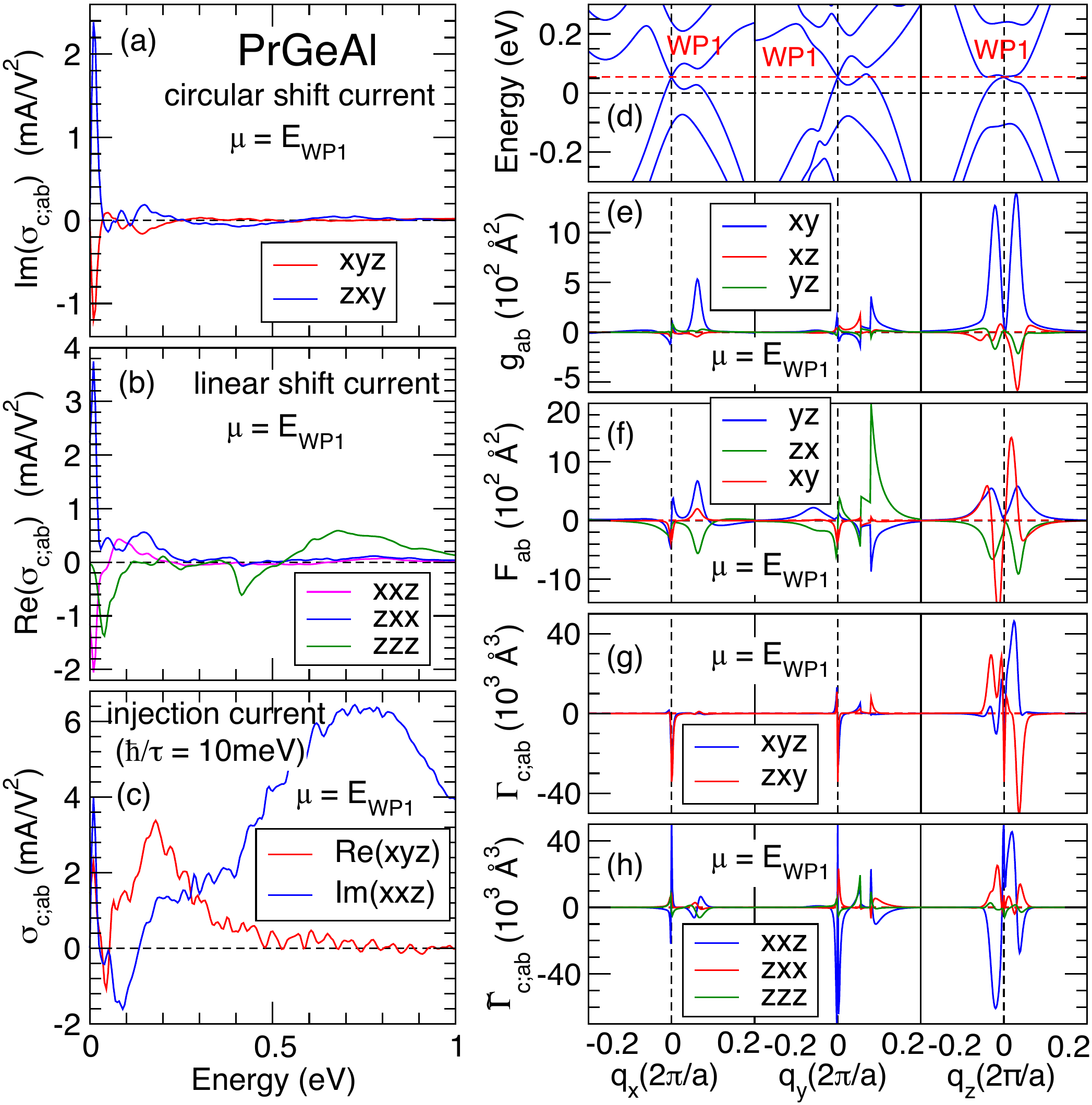}
\caption{
Same as in Fig.~\ref{fig:sigma_PrGeAl}, but with $\mu=E_{\rm WP1}$.
(a, b) Shift and (c) Injection conductivity tensors.
 (d) Energy bands, (e) quantum metric $g_{ab}$, (f) Berry curvature $F_{ab}$,
(g) Christoffel symbol $\Gamma_{c;ab}$, and (h) symplectic Christoffel symbol $\widetilde{\Gamma}_{c;ab}$
along the three cartesian coordinate axis directions through the WP1 Weyl point ($E=E_{\rm WP1}$).
Here ${\bf q}={\bf k}-{\bf k}_{\rm WP1}$ denotes the momentum displacement from the WP1 point.
}
\label{fig:sigmaWP1_PrGeAl}
\end{figure}

The conductivity and geometric quantity spectra obtained by setting the Fermi level to
the WP1 Weyl point energy ($E_{\rm WP1}$) are plotted in Fig.~\ref{fig:sigmaWP1_PrGeAl}.
Remarkably, compared with the $\mu = E_{F}$ case in Fig.~\ref{fig:sigma_PrGeAl}, all the photoconductivity
elements within the photon energy of 50 meV increase by a factor of 2 or more.
In particular, linear shift conductivity element Re($\sigma^{zxx}$)
is enhanced by a factor of 5 [Fig.~\ref{fig:sigmaWP1_PrGeAl}(b)].
Other changes include that the peak at $\sim$70 meV in circular shift conductivity elements
Im($\sigma^{xyz}$) and Im($\sigma^{zxy}$) get significantly reduced [Fig.~\ref{fig:sigmaWP1_PrGeAl}(a)]
and that the positive peak at $\sim$40 meV in linear shift conductivity element Re($\sigma^{zzz}$)
changes to the negative peak [Fig.~\ref{fig:sigmaWP1_PrGeAl}(a)].
Nonetheless, the features of all the conductivity elements above 100 meV
remain essentially unchanged.

The changes caused by moving the Fermi level to the WP1 WP energy mentioned above,
can be largely explained by the distributions of the four geometric quantities
in the vicinity of the WP1 Weyl point, as displayed in Figs. \ref{fig:sigmaWP1_PrGeAl}(e),
\ref{fig:sigmaWP1_PrGeAl}(f), \ref{fig:sigmaWP1_PrGeAl}(g) and \ref{fig:sigmaWP1_PrGeAl}(h).
For example, symplectic Christoffel symbol $\widetilde{\Gamma}_{zxx}$ has high positive
peaks along all three coordinate directions near the WP1 WP [see Figs. \ref{fig:sigmaWP1_PrGeAl}(h)].
These peaks thus give rise to the five-fold increased linear shift conductivity Re($\sigma^{zxx}$).
Note that although in the $\mu = E_{F}$ case $\widetilde{\Gamma}_{zxx}$ also has
sharp peaks near the $\Sigma$ and $\Sigma_1$ points, these peaks have both positive and negative
signs [see Figs. \ref{fig:sigma_PrGeAl}(i)] and thus their contributions to
the linear shift conductivity cancel each other to some extent.
\\

\begin{figure}[t] \centering
\includegraphics[width=8.5cm]{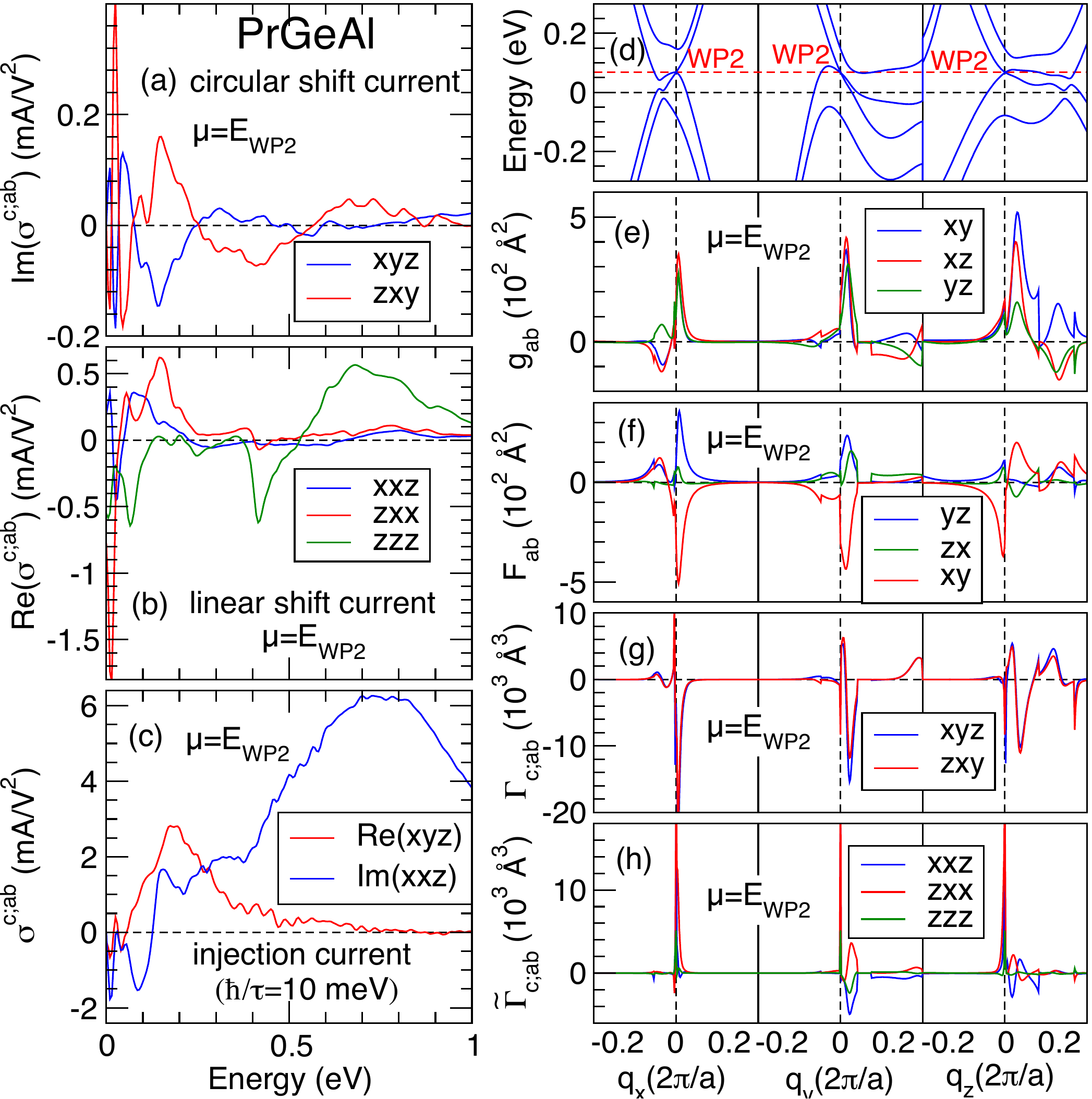}
\caption{
Same as in Fig.~\ref{fig:sigma_PrGeAl}, but with $\mu=E_{\rm WP2}$.
(a, b) Shift and (c) Injection conductivity tensors.
 (d) Energy bands, (e) quantum metric $g_{ab}$, (f) Berry curvature $F_{ab}$,
(g) Christoffel symbol $\Gamma_{c;ab}$, and (h) symplectic Christoffel symbol $\widetilde{\Gamma}_{c;ab}$
along the three cartesian coordinate axis directions through the WP2 Weyl point $E=E_{\rm WP2}$.
${\bf q}={\bf k}-{\bf k}_{\rm WP2}$ denotes the momentum displacement from the WP2 point.
}
\label{fig:sigmaWP2_PrGeAl}
\end{figure}

The results obtained by setting the Fermi level to the energy of the type II WP2 Weyl point
are plotted in Fig.~\ref{fig:sigmaWP2_PrGeAl}.
We notice that as for the WP1 WP case, all the photoconductivity spectra remain more or less
unchanged for photon energies larger than 100 meV.
As for the WP1 WP case, the peak height of linear shift conductivity Re($\sigma^{zxx}$)
at very low energy of $\sim$10 meV gets doubled, although the sign of the peak changes
from positive to negative [see Fig.~\ref{fig:sigmaWP2_PrGeAl}(b)].
However, in contrast to the WP1 Weyl point case, many low energy peaks especially
of shift current conductivity elements, become smaller [Figs. \ref{fig:sigmaWP2_PrGeAl}(a) and \ref{fig:sigmaWP2_PrGeAl}(b)].
In particular, the peak at 10 meV in both circular shift elements Im($\sigma^{xyz}$) and Im($\sigma^{zxy}$)
decreases by a factor of nearly 2, and also the peak sign changes to the opposite.
These different changes in the low energy photoconductivity spectra betwen the type I and type II
Weyl point cases may be attributed to their different energy dispersions and hence the different
distributions of the four geometric quantities near the Weyl points (see Figs. \ref{fig:sigmaWP1_PrGeAl} and \ref{fig:sigmaWP2_PrGeAl}).
Figure \ref{fig:sigmaWP1_PrGeAl}(d) shows that in the type I WP1 case, the upper and lower Weyl cone bands lie, respectively, above
and below the WP1 energy. As a result, the low energy inter-Weyl-band transitions may occurr in all three coordinate directions
when the Fermi level is set to $E_{\rm WP1}$. This can result in large peaks in the four geometric quantities
near the WP1 Weyl point, thus leading to the many fold enhancements in photocurrents, as discussed above.
In the type II WP2 Weyl point case, in contrast, both upper and lower Weyl bands 
along the $-k_x$ and $+k_y$ directions lie below $E_{\rm WP2}$ [see Fig.~\ref{fig:sigmaWP2_PrGeAl}(d)]. 
This hinders the low energy inter-Weyl-band transitions along these two directions
when $E_F$ is set to $E_{\rm WP2}$, and thus reduces the peaks 
in the geometric quantities near the WP2 point, thereby leading to several smaller peaks 
in the photocurrents below $\sim$20 meV.
Of course, in a real topological semimetal such as the present system, situation could be more complicated.
Figure \ref{fig:sigmaWP2_PrGeAl}(d) shows that other bands may come quite close to the WP2 Weyl point 
in energy and they thus enable, e.g., the transitions from two Weyl bands below $E_{\rm WP2}$
to the other bands above $E_{\rm WP2}$ along the $-k_x$ and $+k_y$ directions, thus also giving rise
to prominent peaks in the geometric quantities near the WP2 point
[see Figs. \ref{fig:sigmaWP2_PrGeAl}(e), \ref{fig:sigmaWP2_PrGeAl}(f), \ref{fig:sigmaWP2_PrGeAl}(g)  and \ref{fig:sigmaWP2_PrGeAl}(h)].

\section{Discussion}
\label{sec:discussion}

Let us discuss some issues related to the smallness of the frequency.
We note that our theory is reliable for the photon frequency above 1 THz because we assume
$\omega\gg \Gamma.$
Since the typical relaxation time in semimetals is one picosecond, $\omega/2\pi>1$ THz should be taken for our theory to apply.
Thus, the divergence of the responses in our model should not be interpreted as a physical divergence, and it is cut off at $\omega\sim \Gamma$.
Moreover, intraband (i.e., non-transitive) second-order optical responses exist in magnetic systems, and they become comparable to the interband responses when $\omega\lesssim \Gamma$.
To estimate the magnitude of the intraband responses, let us consider the semiclassical second-order optical response $\sigma^{c;ab}_{\rm SC}(\omega)=-(e^3/\hbar^2)(\omega^2+\Gamma^2)^{-1}\sum_n\int_{\bf k}f^{\rm FD}_{n}\d_a\d_bv^c_{n}$.
It has the same symmetry property as the linear injection current (so it appears in magnetic systems where time-reversal symmetry is broken) and scales as $\omega^{-2}\mu^{2-d}$ near tilted Dirac or Weyl points.
Since the $\omega^{-2}$ factor enhances the semiclassical response at low-frequences, it becomes comparable to the peak value of the linear injection response at $\omega\lesssim \tau^{-1}=\Gamma$ if we take $\mu~\sim \omega$.

Another thing to care about at low frequencies is the validity of the perturbation theory.
We suppose that the photovoltaic responses are dominated by the second-order responses since the electric field is generally weak enough.
However, at small frequencies, $A=E/\omega$ becomes large, so higher-order responses can become significant.
According to the Floquet theory analysis in Ref.~\cite{morimoto2016topological}, perturbation theory works well for
$\left|\omega^{-1}eE v\right|\ll \hbar\Gamma$, where we assume that the interband velocity matrix element has the same order of magnitude as the intraband velocity.
It sets a lower bound for $\omega$.
For $\hbar v\sim 1\; {\rm eV\AA^{-1}}$ and $\Gamma\sim 1$ THz,
the lower bound is $\omega_{\rm min}\approx \sqrt{I/({\rm W/cm^2})} {\rm \;THz}$.
While this bound is very small for typical lasers with intensity $I<1\;{\rm W/cm^2}$, it should be taken into account for an analysis of the experimental results obtained by using high-intensity pulse lasers.

In the non-perturbative regime, light absorption occurs faster than the interband relaxation, so electrons are excited until the system reaches a new equilibrium where the absorption and emission of light finds a balance.
The relaxation plays an important role here because a mismatch between the absorption and emission, which generate opposite photocurrents, requires a finite relaxation~\cite{morimoto2016topological}.
Even in a very clean system with an extremely small $\Gamma$, the low-frequency divergent behavior we discuss above is cut off by this reason.

At finite temperature, thermal fluctuation can significantly reduce low-frequency photovoltaic response in tilted-Dirac or Weyl semimetals~\cite{yang2017divergent}.
When the peak value is considered, the relevant frequency scale where the peak value is reduced to half is $2|v/v'|k_BT/\hbar$ for type I and $2k_BT/\hbar$ for type II.
It sets a quite high lower-bound, since $2k_BT/\hbar=$12 THz at room temperature (cf. 0.17 THz at 4.2K and 3.2 THz at 77 K).
When the tilting is small such that $|v'/v|<0.4$, the cut-off scale is larger than $30\;{\rm THz}$ at room temperature, which is the highest edge of the terahertz radiation.
Therefore, cooling will be needed in order for tilted-Dirac or Weyl semimetals to work as an efficient terahertz photodetector.

Throughout this work, we focus on the DC generation.
However, let us note that our results can also be applied to the second harmonic generation, where an alternating current of frequency $2\omega$ is generated by a uniform illumination of light with frequency $\omega$.
Since the second harmonic generation is associated with the shift vector~\cite{morimoto2016topological,yang2017divergent}, its low-energy divergence has the same form as that for the shift current~\cite{yang2017divergent}, and thus it can be related to the quantum geometric connection.
We believe that similar geometric interpretations of other quantities are also possible.
It will be an interesting topic to explore quantum geometric properties of the third-order optical conductivity.
Since it has four components, we expect the existence of a third-order optical response controlled by the Riemann curvature tensor $R^{a}_{bcd}=\d_c\Gamma^{a}_{db}-\d_d\Gamma^{a}_{cb}+\sum_e(\Gamma^{a}_{ce}\Gamma^{e}_{db}-\Gamma^{a}_{de}\Gamma^{e}_{cb})$.

{\it Note added.---}
Recently, Hikaru Watanabe and Yoichi Yanase studied the bulk photovoltaic responses in magnetic systems independently of ours.
They also figure out the circular shift current and discuss its enhancement near gap-closing points.
Our results are consistent with the results in their preprint when there is an overlap.

\begin{acknowledgments}
We appreciate Takahiro Morimoto, Yoshinori Tokura, Naoki Ogawa, and Bohm-Jung Yang for helpful discussions.
J.A. especially thanks Takahiro Morimoto for participating in the early period of this project and educating much about the shift and injection currents.
We thank Hikaru Watanabe and Yoichi Yanase for communicating their results before submission.
J.A. acknowledges the funding from the RIKEN Special Postdoctoral Researcher Program.
G.-Y. Guo acknowledges the support from the Ministry of Science and Technology and National Center for Theoretical Sciences in Taiwan.
G.-Y. Guo also thanks the National Center for High-performance Computing in Taiwan for the computing time.
This work was supported by JST CREST Grant Number JPMJCR1874 and JPMJCR16F1, Japan, and JSPS KAKENHI Grant Numbers 18H03676 and 26103006.
\end{acknowledgments}

\appendix

\section{Conventions and Definitions}

\begin{itemize}
\item We do not distinguish uppercase and lowercase indices except for the Christoffel symbols.

\item $-e$ is the electron charge. $e>0$.

\item Geometric quantities on the Bloch sphere (indices $i=1,2,3$ are for the $(f_1,f_2,f_3)$ basis):

\begin{align}
&\eta_{ij}									&\text{metric tensor}\notag\\
&\epsilon_{ij}								&\text{symplectic form}\notag\\
&q_{ij}=\eta_{ij}-i\epsilon_{ij}					&\text{quantum geometric tensor}\notag\\
&\gamma^{i}_{jk} 								&\text{Christoffel symbol of the second kind}\notag\\
&\gamma_{ijk} 								&\text{Christoffel symbol of the first kind}\notag\\
&\tilde{\gamma}_{ijk} 						&\text{symplectic Christoffel symbol}\notag\\
&c_{ijk}=\gamma_{ijk}-i\tilde{\gamma}_{ijk} 	&\text{quantum geometric connection}
\end{align}

\item
Geometric quantities in momentum space (indices $a=1,2,3$ are for the $(k_1,k_2,k_3)$ basis)
\begin{align}
&g_{ab}										&\text{metric tensor}\notag\\
&F_{ab}/2									&\text{symplectic form}\notag\\
&Q_{ab}=g_{ab}-iF_{ab}/2						&\text{quantum geometric tensor}\notag\\
&\Gamma^{a}_{bc} 								&\text{Christoffel symbol of the second kind}\notag\\
&\Gamma_{abc} 								&\text{Christoffel symbol of the first kind}\notag\\
&\tilde{\Gamma}_{abc} 						&\text{symplectic Christoffel symbol}\notag\\
&C_{abc}=\Gamma_{abc}-i\tilde{\Gamma}_{abc}&\text{quantum geometric connection}
\end{align}

\item
Typical quantities in second-order optical conductivity.
\begin{align}
\d_a
&=\frac{\d}{\d_{k_a}},\notag\\
\int_{\bf k}
&=\int \frac{d^3k}{(2\pi)^3},\notag\\
\omega_n
&=\hbar^{-1}\braket{n|H|n},\notag\\
v^a_{mn}
&=\hbar^{-1}\braket{m|\d_aH|n},\notag\\
w^{ab}_{mn}
&=\hbar^{-1}\braket{m|\d_a\d_bH|n},\notag\\
u^{abc}_{mn}
&=\hbar^{-1}\braket{m|\d_a\d_b\d_cH|n},\notag\\
f^{\rm FD}_{m}
&=\text{Fermi-Dirac distribution of the $m$th band},\notag\\
\omega_{mn}
&=\omega_m-\omega_n,\notag\\
f^{\rm FD}_{mn}
&=f^{\rm FD}_m-f^{\rm FD}_n,\notag\\
\Delta^a_{mn}
&=v^a_{mm}-v^a_{nn},\notag\\
R^{c;a}_{mn}
&=r^c_{mm}-r^c_{nn}+i\d_c\log r^a_{mn},\notag\\
r^a_{mn}
&=i\braket{m|\d_a|n},\notag\\
r^a_{mn;c}
&=\d_cr^a_{mn}-i(r^c_{mm}-r^c_{nn})r^a_{mn}.
\end{align}
\end{itemize}

\section{Frequently used identities}

In the appendix we use the following two identities.
\begin{align}
\label{eq:identity}
\braket{m|\d_aO|n}
&=\d_aO_{mn}
-i(r^a_{mp}O_{pn}-O_{mp}r^a_{pn})\notag\\
&=\d_aO_{mn}
-i[r^a,O]_{mn},
\end{align}
where we integrate by parts in the first line.
The following is a specific example of the above identity with $O=H/\hbar$, and it is used very often in the following appendices.
\begin{align}
\label{eq:identity2}
v^a_{mm}
&=\d_a\omega_{m}\notag\\
v^a_{mn}
&=i\omega_{mn}r^a_{mn}\quad (m\ne n).
\end{align}

\section{Shift and injection currents from the Fermi Golden rule}
\label{sec:Golden}

Let us derive the expression of the shift and injection currents from the Fermi Golden rule.
Here, we drop the light polarization dependence of the shift vector.
Since ${\bf E}(t)=E(\omega)e^{-i\omega t}+E(-\omega)e^{i\omega t}$, where $E(-\omega)=E(\omega)^*$, both $\omega$ and $-\omega$ components contribute.
\begin{align}
&j^{c}_{\rm shift}\notag\\
&=
\int_{\bf k}
\sum_{\substack{n\in{\rm occ}\\m\in{\rm unocc}}}
(-eR^c_{mn})
f^{\rm FD}_{nm}{\cal M}_{m\leftarrow n}\notag\\
&=
\bigg[\int_{\bf k}
\sum_{\substack{n\in{\rm occ}\\m\in{\rm unocc}}}
(-eR^c_{mn})f^{\rm FD}_{nm}\frac{2\pi}{\hbar}\left|\sum_a\braket{m|\frac{ie}{\hbar \omega}E_a(\omega)\d_aH|n}\right|^2\notag\\
&\qquad \times \delta(\hbar\omega_{mn}-\hbar\omega)\bigg]
+(\omega\rightarrow -\omega)\notag\\
&=-
\frac{2\pi e^3}{\hbar}
\int_{\bf k}
\sum_{n,m,a,b}
f^{\rm FD}_{nm}R^c_{mn} \frac{\braket{n|\d_bH|m}}{\hbar\omega_{mn}}
\frac{\braket{m|\d_aH|n}}{\hbar\omega_{mn}}\notag\\
&\qquad \times E_a(\omega)E_b(-\omega)\delta(\hbar\omega_{mn}-\hbar\omega)\notag\\
&=-
\frac{2\pi e^3}{\hbar^2}
\int_{\bf k}
\sum_{n,m,a,b}
f^{\rm FD}_{nm}R^c_{mn}
r^b_{nm}r^a_{mn}
\delta(\omega_{mn}-\omega)\notag\\
&\qquad \times E_a(\omega)E_b(-\omega)\notag\\
&\d_tj^{c}_{\rm inj}\notag\\
&=
\int_{\bf k}
\sum_{n,m,a,b}
(-e\Delta^c_{mn})
f^{\rm FD}_{nm}{\cal M}_{m\leftarrow n}(\omega)\notag\\
&=-
\frac{2\pi e^3}{\hbar^2}
\int_{\bf k}
\sum_{n,m}
f^{\rm FD}_{nm}\Delta^c_{mn}
r^b_{nm}r^a_{mn}
\delta(\omega_{mn}-\omega)\notag\\
&\qquad \times E_a(\omega)E_b(-\omega),
\end{align}
where ${\cal M}_{m\leftarrow n}$ is the probability of the transition from $n$ to $m$.
We note that this Fermi Golden rule calculation gives an overall $(-1)$ sign that is absent in the result of in~\cite{parker2019diagrammatic,holder2020consequences} while they also take $-e$ as the electron charge.
Also, our expressions differ from the result in~\cite{parker2019diagrammatic,holder2020consequences} by the sign of $\omega$ in the electric fields.

\section{Time reversal transformation of the second-order optical conductivity}
\label{sec:time}

Let us consider the most general second-order optical conductivity tensor defined by
\begin{align}
\label{eq:general-second}
j^c(\tilde{\omega})=\sigma^{c;ab}(\tilde{\omega};\tilde{\omega}_1,\tilde{\omega}_2)E_a(\tilde{\omega}_1)E_b(\tilde{\omega}_2),
\end{align}
where $\tilde{\omega}=\tilde{\omega}_1+\tilde{\omega}_2$, and $\tilde{\omega}_{1,2}=\omega_{1,2}+i\Gamma$.
We can symmetrize the conductivity without loss of generality such that
\begin{align}
\sigma^{c;ab}(\tilde{\omega};\tilde{\omega}_1,\tilde{\omega}_2)
=\sigma^{c;ba}(\tilde{\omega};\tilde{\omega}_2,\tilde{\omega}_1).
\end{align}
The current and the electric field transform as
\begin{align}
j^c(\tilde{\omega})
&\rightarrow j'^c(\tilde{\omega})=-j^c(-\tilde{\omega})\notag\\
E_a(\tilde{\omega}_1)
&\rightarrow E'_a(\tilde{\omega}_1)=E_a(-\tilde{\omega}_1)
\end{align}
under time reversal $t\rightarrow -t$.
Equation~\eqref{eq:general-second} is equivalent to
$-j'^c(-\tilde{\omega})=\sigma^{c;ab}(\tilde{\omega};\tilde{\omega}_1,\tilde{\omega}_2)E'_a(-\tilde{\omega}_1)E'_b(-\tilde{\omega}_2).$
It can be written as 
\begin{align}
j'^c(\tilde{\omega})=\sigma'^{c;ab}(\tilde{\omega};\tilde{\omega}_1,\tilde{\omega}_2)E'_a(\tilde{\omega}_1)E'_b(\tilde{\omega}_2)
\end{align}
if we define $\sigma'^{c;ab}$ by
\begin{align}
\sigma'^{c;ab}(\tilde{\omega};\tilde{\omega}_1,\tilde{\omega}_2)
&=-\sigma^{c;ab}(-\tilde{\omega};-\tilde{\omega}_1,-\tilde{\omega}_2)\notag\\
&=-\sigma^{c;ba}(-\tilde{\omega};-\tilde{\omega}_2,-\tilde{\omega}_1).
\end{align}
In the case of the DC generation where $\omega_1=-\omega_2$, this reduces to
\begin{align}
\label{eq:time-reversal}
\sigma'^{c;ab}(\omega,\Gamma)
=-\sigma^{c;ba}(\omega,-\Gamma).
\end{align}
where $\sigma^{c;ab}(\omega,\Gamma)=\sigma^{c;ab}(0;\omega+i\Gamma,-\omega+i\Gamma)$.

\section{Derivaion of Eqs.~\eqref{eq:linear-transform} and~\eqref{eq:circular-transform}}
\label{sec:transformation}

Let ${\cal M}$ be a point-group symmetry operation: $x_a\rightarrow x'_a={\cal M}_{ab}x_b$.
The Bloch state transforms as
\begin{align}
\ket{\psi_{n\bf k}}
\rightarrow \ket{\psi'_{n\bf k}}
=\hat{\cal M}\ket{\psi_{n{\cal M}^{-1}\bf k}}.
\end{align}
When the system has ${\cal M}$ symmetry (i.e., $\hat{\cal M}^{-1}H\hat{\cal M}=H$), $\ket{\psi'_{n\bf k}}=\ket{\psi_{m\bf k}}M_{mn}({\bf k})$ is satisfied, where $M_{mn}$({\bf k}) is a unitary matrix that is nonzero only when $E_{m}({\bf k})=E_n({\cal M}^{-1}{\bf k})$.
Here, we {\it do not} assume that the system has $\cal M$ symmetry and focus on how quantities transform under $\cal M$.
From the transformation property of the Bloch state, it follows that
\begin{align}
\ket{u_{n\bf k}}
\rightarrow \ket{u'_{n\bf k}}
=\hat{\cal M}\ket{u_{n{\cal M}^{-1}\bf k}}.
\end{align}
Then,
\begin{align}
r^a_{mn}({\bf k})\rightarrow r^{a'}_{mn}({\bf k})=g_{ab}r^b_{mn}({\cal M}^{-1}{\bf k}).
\end{align}
We can see this from
\begin{align}
r'^{a}_{mn}({\bf k})
&\equiv \braket{u'_{m\bf k}|i\d_a|u'_{n\bf k}}\notag\\
&=\braket{\hat{\cal M}u_{m{\cal M}^{-1}\bf k}|i\d_a|\hat{\cal M}u'_{n{\cal M}^{-1}\bf k}}\notag\\
&=\frac{\d({\cal M}^{-1}k)_b}{\d k_a}\braket{\hat{\cal M}u_{m{\cal M}^{-1}\bf k}|i\d_{({\cal M}^{-1}k)_b}|\hat{\cal M}u'_{n{\cal M}^{-1}\bf k}}\notag\\
&={\cal M}_{ab}r^b_{mn}({\cal M}^{-1}{\bf k}),
\end{align}
where we use that $({\cal M}^{-1})_{ba}={\cal M}_{ab}$.
Similarly, we have
\begin{align}
&v'^{a}_{mn}({\bf k})\notag\\
&\equiv \hbar^{-1}\braket{u'_{m\bf k}|\d_aH'({\bf k})|u'_{n\bf k}}\notag\\
&=\hbar^{-1}\braket{\hat{\cal M}u_{m{\cal M}^{-1}\bf k}|\d_a(\hat{\cal M}H({\cal M}^{-1}{\bf k})\hat{\cal M}^{-1})|\hat{\cal M}u'_{n{\cal M}^{-1}\bf k}}\notag\\
&={\cal M}_{ab}v^b_{mn}({\cal M}^{-1}{\bf k}),
\end{align}
where $H'({\bf k})=\hat{\cal M}H({\bf k})\hat{\cal M}^{-1}$, and $H({\bf k})=e^{-i{\bf k}\cdot\hat{\bf r}}He^{i{\bf k}\cdot\hat{\bf r}}$ is the Bloch Hamiltonian.
From these properties, one can find that
\begin{align}
\sigma'^{c_1;a_1b_1}_{\rm DC}(\omega,\Gamma)
={\cal M}_{c_1c}{\cal M}_{a_1a}{\cal M}_{b_1b}\sigma^{c;ab}_{\rm DC}(\omega,\Gamma),
\end{align}
where $\sigma'^{c;ab}_{\rm DC}$ is defined from the ${\cal M}$-transformed state $\ket{u'_{n\bf k}}$ and Hamiltonian $H'({\bf k})$.

We can repeat this process for time reversal
\begin{align}
\ket{u_{n\bf k}}
\rightarrow \ket{\psi'_{n\bf k}}
&=\hat{T}\ket{u_{n, -\bf k}}.
\end{align}
For the dipole matrix elements,
\begin{align}
r'^{a}_{mn}({\bf k})
&\equiv \braket{u'_{m\bf k}|i\d_a|u'_{n\bf k}}\notag\\
&=\braket{\hat{T}u_{m,-\bf k}|i\d_a|\hat{T}u'_{n,-\bf k}}\notag\\
&=\braket{\hat{T}u_{m,-\bf k}|\hat{T}(-i)\d_au'_{n,-\bf k}}\notag\\
&=\braket{(-i)\d_au'_{n,-\bf k}|u_{m,-\bf k}}\notag\\
&=\braket{u'_{n,-\bf k}|i\d_a|u_{m,-\bf k}}\notag\\
&=r^a_{nm}(-{\bf k}),
\end{align}
where we use the antiunitary property of time reversal in the third line, and we integrate by parts in the fourth line.
\begin{align}
v'^{a}_{mn}({\bf k})
&\equiv \hbar^{-1}\braket{u'_{m\bf k}|\d_aH'({\bf k})|u'_{n\bf k}}\notag\\
&=\hbar^{-1}\braket{\hat{T}u_{m,-\bf k}|\d_a(\hat{T}H(-{\bf k})\hat{T}^{-1})|\hat{T}u'_{n,-\bf k}}\notag\\
&=\hbar^{-1}\braket{\hat{T}u_{m,-\bf k}|\hat{T}\d_a[H(-{\bf k})]u'_{n,-\bf k}}\notag\\
&=\hbar^{-1}\braket{\d_a[H(-{\bf k})]u'_{n,-\bf k}|u_{m,-\bf k}}\notag\\
&=-\hbar^{-1}\braket{(\d_aH)(-{\bf k})u'_{n,-\bf k}|u_{m,-\bf k}}\notag\\
&=-v^b_{nm}(-{\bf k}),
\end{align}
We have
\begin{align}
\sigma'^{c;ab}_{\rm DC}(\omega,\Gamma)
=\sigma^{c;ab}_{\rm DC}(\omega,\Gamma),
\end{align}
for linear shift and circular injection, and
\begin{align}
\sigma'^{c;ab}_{\rm DC}(\omega,\Gamma)
=-\sigma^{c;ab}_{\rm DC}(\omega,\Gamma),
\end{align}
for circular shift and linear injection under time reversal.
This proves Eqs.~\eqref{eq:linear-transform} and~\eqref{eq:circular-transform}.
Let us note that the time-reversal transformations derived here are consistent with Eq.~\eqref{eq:time-reversal}.

\section{General form of shift current matrix elements}

In the clean limit, the DC conductivity corresponding to the shift current is
\begin{align}
\sigma^{c;ab}_{\rm shift}
&=\frac{\pi e^3}{\hbar^2}
\sum_{n,m}
\int_{\bf k}
f^{\rm FD}_{nm}
I^{c;ab}_{nm}
\delta(\omega_{mn}-\omega),
\end{align}
where
\begin{align}
I^{c;ab}_{nm}=-i(r^b_{nm}r^a_{mn,c}-r^b_{nm,c}r^a_{mn}),
\end{align}
and
\begin{align}
r^a_{mn,c}
\equiv \d_cr^a_{mn}-i(r^c_{mm}-r^c_{nn})r^a_{mn}.
\end{align}
One can see that $I^{c;ab}_{nm}$ is identical to $(R^{c;a}_{mn}-R^{c;b}_{nm})r^b_{nm}r^a_{mn}$.
Let us transform this to an expression involving only matrix elements of the derivatives of the Hamiltonian.
Here it is enough to consider $n\in \rm occ$ and $m\in \rm unocc$ because of the Fermi-Dirac distribution function.
We first use Eq.~\eqref{eq:identity2} to write the derivative of $r^a_{mn}$ in terms of the velocity matrix elements.
\begin{align}
\d_cr^a_{mn}
&=\d_c\left(\frac{ v^a_{mn}}{i \omega_{mn}}\right)\notag\\
&=\frac{1}{i \omega_{mn}}
\left[ \d_cv^a_{mn}- v^a_{mn}\frac{\d_c\omega_{mn}}{\omega_{mn}}\right].
\end{align}
The derivative of the velocity matrix element can be decomposed further as
\begin{align}
\d_cv^a_{mn}
&=\hbar^{-1}\braket{m|\d_a\d_cH|n}
+i\hbar^{-1}[r^c,\d_aH]_{mn}\notag\\
&=
w^{ac}_{mn}+i r^c_{mm}v^a_{mn}-i v^a_{mm}r^c_{mn}\notag\\
&\quad+i r^c_{mn}v^a_{nn}-i v^a_{mn}r^c_{nn}\notag\\
&\quad+i \sum_{p\ne m,n}\left(r^c_{mp}v^a_{pn}-v^a_{mp}r^c_{pn}\right)\notag\\
&=
w^{ac}_{mn}
+i(r^c_{mm}-r^c_{nn}) v^a_{mn}
-ir^c_{mn}\Delta^a_{mn}\notag\\
&\quad+\sum_{p\ne m,n}\left(\frac{v^c_{mp}v^a_{pn}}{\omega_{mp}}-\frac{v^a_{mp}v^c_{pn}}{\omega_{pn}}\right),
\end{align}
where we use Eq.~\eqref{eq:identity} and separate $p=m$ or $p=n$ components from the $p\ne m,n$ components in the summation in the second line, and we use Eq.~\eqref{eq:identity} again in the last line.
Using the notation $\d_c \omega_{nm}=v^c_{nn}-v^c_{mm}=\Delta^c_{nm}$, we have
\begin{align}
\label{eq:rbmnc}
r^a_{mn,c}
=\frac{1}{i\omega_{mn}}
\bigg[
&w^{ac}_{mn}
-\frac{ v^c_{mn}\Delta^a_{mn}
+ v^a_{mn}\Delta^c_{mn}}
{\omega_{mn}}\notag\\
&+\sum_{p\ne m,n}\left(\frac{v^c_{mp}v^a_{pn}}{\omega_{mp}}-\frac{v^a_{mp}v^c_{pn}}{\omega_{pn}}\right)
\bigg].
\end{align}
Accordingly, we have
\begin{align}
\label{eq:shift-matrix}
r^b_{nm}r^a_{mn,c}
=
&\frac{ v^b_{nm}}{\omega_{mn}^2}
\left[
w^{ac}_{mn}
-\frac{
 v^c_{mn}\Delta^a_{mn}
+ v^a_{mn}\Delta^c_{mn}}
{\omega_{mn}}
\right]\notag\\
&+\frac{v^b_{nm}}{\omega^2_{mn}}
\sum_{p\ne m,n}\left(\frac{v^c_{mp}v^a_{pn}}{\omega_{mp}}-\frac{v^a_{mp}v^c_{pn}}{\omega_{pn}}\right).
\end{align}

Let us note that the last term in Eq.~\eqref{eq:shift-matrix} (the virtual transitions) in fact should be summed over nondegenerate transitions, i.e., $\sum_{p\ne n,m}$ should be $\sum_{\omega_p\ne \omega_n,\omega_m}$.
To see this, we note that the 3-band term is (which is the most clear from diagrammatic calculations with substitution $\omega\rightarrow \omega+i\Gamma$ in the calculation of $\sigma^{c;ab}(\omega;\omega_1,\omega_2$)~\cite{parker2019diagrammatic,holder2020consequences})
\begin{align}
\frac{v^b_{nm}}{\omega^2_{mn}}
\sum_{p\ne m,n}\frac{v^c_{mp}v^a_{pn}}{\omega_{mp}+i\Gamma}-\frac{v^a_{mp}v^c_{pn}}{\omega_{pn}+i\Gamma}.
\end{align}
Applying $r^a_{pn}=v^a_{pn}/i\omega_{pn}$ (and also $r^a_{mp}=v^a_{mp}/i\omega_{mp}$) gives Eq.~\eqref{eq:shift-matrix} assuming nondegeneracy of bands and taking $\Gamma\rightarrow 0$.
However, in the degenerate case where $\omega_{p}=\omega_{n}$ for $p\ne n$, $v^a_{pn}=i\omega_{pn}r^a_{pn}$ does not imply $r^a_{pn}=v^a_{pn}/i\omega_{pn}$.
Those degenerate cases should be omitted in the summation in Eq.~\eqref{eq:shift-matrix} because $v_{mp} v_{pn}\propto \omega_{pm}\omega_{np}=0$ when degeneracy occurs.

\section{Second-order response of general ${\bf k}$-linear Dirac Hamiltonian}
\label{sec:k-linear}

Let us show that the following Hamiltonian have vanishing shift current and linear injection current responses.
\begin{align}
H({\bf k})
=\hbar \sum_{a,i=1}^dv_{ai}k_a\Gamma_i.
\end{align}
Since this Hamiltonian is linear in momentum, the diamagnetic term $w^{ac}_{mn}$ in Eq.~\eqref{eq:shift-matrix} vanishes.
It follows that the integrand of the second-order optical conductivity is a tensor under the map $f({\bf k})$ where $f^i=\hbar\sum_{a}v_{ai}k_a$, i.e.,
\begin{align}
\sigma^{c;ab}_{\rm DC}
&=\int d^dk \d_cf^k\d_af^i\d_bf^j T_{ijk}.
\end{align}
where $T_{ijk}$ is a tensor in $f$-space.
\begin{align}
\sigma^{c;ab}_{\rm DC}
&=\int d^df \det\left(\frac{\d f_i}{\d k_a}\right)^{-1}\d_cf^k\d_af^i\d_bf^j T_{kij}\notag\\
&=\hbar^3 v_{ai}v_{bj}v_{ck}\det(\hbar v_{ai})\int d^dfT_{kij}({\bf f})\notag\\
&=\hbar^6v_{ai}v_{bj}v_{ck}\det(v_{ai})\int d^dkT_{kij}({\bf k})\notag\\
&=\hbar^6v_{ai}v_{bj}v_{ck}\det(v_{ai})\sigma^{k;ij}_{\rm DC,0},
\end{align}
where we changed the name of the variable from ${\bf f}$ to ${\bf k}$ in the third line, and $\sigma^{k;ij}_0=\int d^dkT_{kij}({\bf k})$ is the conductivity for $\hbar v_{ai}=\delta_{ai}$.
Since the shift current and linear injection current part of $\sigma^{k;ij}_{\rm DC,0}$ vanishes, this finishes the proof.

\section{Quantum geometric tensor for Dirac Hamiltonians}
\label{sec:QGT}
\subsection{Quantum geometric tensor}

The quantum geometric tensor of the occupid states is defined by
\begin{align}
Q_{ab}
&=\sum_{n\in {\rm occ}}\sum_{m\in {\rm unocc}}r^a_{nm}r^b_{mn},
\end{align}
where $r^a_{nm}=i\braket{n|\d_a|m}$.
Its symmetric and antisymmetric parts respectively correspond to the quantum metric and the Berry curvature of the occupied states.

To see this, let us take occupied bands $n_1$ and $n_2$.
Then,
\begin{align}
&\sum_{m\in {\rm unocc}}r^a_{n_1m}r^b_{mn_2}\notag\\
&=\sum_{m\in {\rm unocc}}\braket{n_1|i\d_a |m}\braket{m|i\d_b |n_2}\notag\\
&=\sum_{m\in {\rm unocc}}\braket{\d_a n_1|m}\braket{m|\d_b n_2}\notag\\
&=\braket{\d_an_1| \d_bn_2}
-\sum_{p\in {\rm occ}}\braket{\d_an_1 |p}\braket{p|\d_b n_2}\notag\\
&=
\frac{1}{2}\left[\braket{\d_an_1| \d_bn_2}
-\sum_{p\in {\rm occ}}\braket{\d_an_1 |p}\braket{p|\d_b n_2}+(a\leftrightarrow b)\right]\notag\\
&+
\frac{1}{2}\left[\braket{\d_an_1| \d_bn_2}
-\sum_{p\in {\rm occ}}\braket{\d_an_1 |p}\braket{p|\d_b n_2}-(a\leftrightarrow b)\right]
\notag\\
&=(g_{ab})_{n_1n_2}-\frac{i}{2}(F_{ab})_{n_1n_2},
\end{align}
where the symmetric part
\begin{align}
(g_{ab})_{n_1n_2}
=
&\frac{1}{2}\left[\braket{\d_a n_1|\d_b n_2}
-\sum_{p\in {\rm occ}}\braket{\d_a n_1| p}\braket{p|\d_b n_2}\right]\notag\\
&+(a\leftrightarrow b)
\end{align}
is the nonabelian quantum metric of the occupied states, and the antisymmetric part
\begin{align}
(F_{ab})_{n_1n_2}
&=i\braket{\d_a n_1|\d_b n_2}
-i\sum_{p\in {\rm occ}}\braket{\d_a n_1| p}\braket{p|\d_b n_2}\notag\\
&\quad -(a\leftrightarrow b)\notag\\
&=\d_ar^b_{n_1n_2}-\d_br^a_{n_1n_2}
-i(r^a_{n_1p}r^b_{pn_2}-r^b_{n_1p}r^a_{pn_2})
\end{align}
is the nonabelian Berry curvature of the occupied states, where $r^a_{n_1n_2}=i\braket{n_1|\d_a|n_2}$ is the nonabelian Berry connection of the occupied states.
It follows that
\begin{align}
Q_{ab}
&=\sum_{n\in {\rm occ}}(g_{ab})_{nn}-\frac{i}{2}(F_{ab})_{nn}\notag\\
&=g_{ab}-\frac{i}{2}F_{ab}.
\end{align}

\subsection{Dirac Hamiltonian}
\label{sec:QGT-Dirac}

Here, we consider the following $d_M\times d_M$ Dirac Hamiltonian
\begin{align}
H
&=
f_0+\sum_if_i\Gamma_i.
\end{align}
Let us note that this kind of Hamiltonian describe genetric two-band systems when $d_M=2$ and generic $PT$-symmetric four-band systems with $(PT)^2=-1$ when $d_M=4$.
Let us suppose that half of the bands are occupied.
Then, since $\hbar\omega_{\rm unocc}-\hbar\omega_{\rm occ}=2\sqrt{\sum_if_i^2}=2f$, we have
\begin{align}
Q_{ab}
&=
\frac{\hbar^2}{4f^2}
\sum_{n\in{\rm occ}}
\sum_{m\in {\rm unocc}}
v^a_{nm}v^b_{mn}.
\end{align}
Also, the velocity matrix elements are given by
\begin{align}
\hbar v^a_{m\ne n}
&=\braket{m|\d_aH|n}
=\sum_i\d_af_i\cdot\braket{m|\Gamma_i|n}.
\end{align}
It follows that
\begin{align}
Q_{ab}
&=
\frac{1}{4f^2}
\sum_{i,j}
\d_af^i\d_bf^j
\sum_{n\in{\rm occ}}
\sum_{m\in {\rm unocc}}
\braket{n|\Gamma_i|m}\braket{m|\Gamma_j|n}.
\end{align}
Let us evaluate the summation
\begin{align}
&\sum_{n\in{\rm occ}}
\sum_{m\in {\rm unocc}}
\braket{n|\Gamma_i|m}\braket{m|\Gamma_j|n}\notag\\
&=
\sum_{n\in{\rm occ}}
\left[\braket{n|\Gamma_i\Gamma_j|n}
-
\sum_{m\in {\rm occ}}
\braket{n|\Gamma_i|m}\braket{m|\Gamma_j|n}\right].
\end{align}
The first term is
\begin{align}
\label{eq:Gamma-Gamma}
\sum_{n\in {\rm occ}}\braket{n|\Gamma_i\Gamma_j|n}
&=
\sum_{n\in {\rm occ}}\frac{1}{2}\braket{n|\{\Gamma_i,\Gamma_j\}|n}
+\frac{1}{2}\braket{n|[\Gamma_i,\Gamma_j]|n}\notag\\
&=
\frac{d_M}{2}(\delta_{ij}+iJ_{ij}),
\end{align}
where we use the property of Gamma matrices $\{\Gamma_i,\Gamma_j\}=2\delta_{ij}$ and define
\begin{align}
J_{ij}
\equiv \frac{i}{d_M}\sum_{n\in {\rm occ}}\braket{n|[\Gamma_i,\Gamma_j]|n}.
\end{align}
As for the second term, we note that for $m,n\in {\rm occ}$,
\begin{align}
\label{eq:Gamma_mn}
\braket{m|\Gamma_j|n}
=-\frac{f_j}{f}\delta_{mn}
=-\hat{f}_j\delta_{mn}.
\end{align}
One can see this as follows.
First, if we dot product $\braket{m|\Gamma_j|n}$ with ${\bf f}$, we have $\braket{m|\Gamma_j|n}f_j=\braket{m|f_j\Gamma_j|n}=-f\delta_{mn}$.
On the other hand, for a vector ${\bf n}$ perpendicular to ${\bf f}$, $\braket{m|\Gamma_j|n}n_j=\braket{m|n_j\Gamma_j|n}=0$ because $n_j\Gamma_j$ anticommutes with $f_i\Gamma_i$.
Since $\braket{m|\Gamma_j|n}$ is parallel to $f_i$, we have Eq.~\eqref{eq:Gamma_mn}.
To sum up, we have
\begin{align}
\label{eq:quantum-metric}
Q_{ab}
&=
\sum_{i,j}
\d_af^i\d_bf^j
\frac{d_M(\delta_{ij}-\hat{f}_i\hat{f}_j+iJ_{ij})}{8f^2}.
\end{align}
In two-band systems, we have $J_{ij}=-\epsilon_{ijk}\hat{f}_k$ because
$[\Gamma_i,\Gamma_j]=2i\epsilon_{ijk}\Gamma_k$, where $\Gamma_i=\sigma_i$ is a Pauli matrix.
On the other hand, in four-band systems described by a Dirac Hamiltonian, which are $PT$-symmetric four-band systems, we have $J_{ij}=0$.

\section{Compatibility between metric, symplectic form, and connection}

\subsection{Metric connection}

A connection $\gamma^k_{ij}$ is called metric compatible if the inner product done by the metric is invariant under parallel transport~\cite{nakahara2003geometry}:
\begin{align}
\nabla\eta
=0.
\end{align}
In components, it reads
\begin{align}
\d_k\eta_{ij}
-\eta_{il}\gamma^l_{kj}
-\eta_{jl}\gamma^l_{ij}=0.
\end{align}
This can be seen like this.
\begin{align}
0
&=(\nabla_k\eta)_{ij}\notag\\
&={\bf e}_i\cdot \nabla_k\eta\cdot {\bf e}_j\notag\\
&=\nabla_k({\bf e}_i\cdot \eta\cdot {\bf e}_j)
-\nabla_k{\bf e}_i\cdot\eta\cdot {\bf e}_j
-{\bf e}_i\cdot\eta\cdot \nabla_k{\bf e}_j\notag\\
&=\d_k\eta_{ij}-\gamma^l_{ki}\eta_{lj}-\eta_{il}\gamma^l_{kj},
\end{align}
where we use the notation $\eta_{ij}={\bf e}_i\cdot \eta\cdot {\bf e}_j$ and the definition $\nabla_k{\bf e}_i=\gamma^l_{ki}{\bf e}_l$.
A metric compatible connection is called a metric connection.
If a metric connection $\gamma^k_{ij}$ satisfies the torsion-free condition $\gamma^k_{ij}=\gamma^k_{ji}$ (torsion is the antisymmetric part of $\gamma^k_{ij}$), it is uniquely determined to be
\begin{align}
\gamma^{k}_{ij}
&=g^{kl}\frac{1}{2}\left(\d_i\eta_{lj}+\d_{j}\eta_{li}-\d_l\eta_{ij}\right),
\end{align}
and it is called the Levi-Civita connection.

\subsection{Symplectic connection}

Similarly, a connection is called a symplectic connection when it satisfies
\begin{align}
\nabla \epsilon
=0.
\end{align}
In components, it reads
\begin{align}
\d_k\epsilon_{ij}
-\epsilon_{il}\gamma^l_{kj}
+\epsilon_{jl}\gamma^l_{ki}=0.
\end{align}
This follows from
\begin{align}
0
&=(\nabla_k\epsilon)_{ij}\notag\\
&={\bf e}_i\cdot \nabla_k\epsilon\cdot {\bf e}_j\notag\\
&=\nabla_k({\bf e}_i\cdot \epsilon\cdot {\bf e}_j)
-\nabla_k{\bf e}_i\cdot\epsilon\cdot {\bf e}_j
-{\bf e}_i\cdot\epsilon\cdot \nabla_k{\bf e}_j\notag\\
&=\d_k\epsilon_{ij}-\gamma^l_{ki}\epsilon_{lj}-\epsilon_{il}\gamma^l_{kj}\notag\\
&=\d_k\epsilon_{ij}+\gamma^l_{ki}\epsilon_{jl}-\epsilon_{il}\gamma^l_{kj}.
\end{align}
Unlike the metric compatibility, this condition alone cannot uniquely determine the connection even after imposing the torsion-free condition $\gamma^k_{ij}=\gamma^k_{ji}$.
However, adding metric compatibility makes the connection unique.
On the generalized Bloch sphere, the Levi-Civita connection is the unique connection that is compatible with both the metric and the symplectic form.

\begin{widetext}

\section{Geometric interpretation of the circular shift current with more than two bands}
\label{sec:multiband}

Here we show that the matrix element of the circular shift conductivity is related with the Christoffel symbol of the first kind defined between the $n$ and $m$ bands:
\begin{align}
\Gamma_{bca;nm}
=\frac{1}{2}\left(\d_cg_{ba;nm}+\d_ag_{bc;nm}-\d_bg_{ca;nm}\right),
\end{align}
where we define
\begin{align}
g_{ab;nm}
&={\rm Re}\left(r^a_{nm}r^b_{mn}\right).
\end{align}
By using the identity
\begin{align}
r^b_{nm}r^a_{mn,c}+r^b_{nm,c}r^a_{mn}
&=
r^b_{nm}\d_cr^a_{mn}-ir^b_{nm}r^a_{mn}(r^c_{mm}-r^c_{nn})
+
(\d_cr^b_{nm})r^a_{mn}-ir^b_{nm}r^a_{mn}(r^c_{nn}-r^c_{mm})\notag\\
&=\d_c(r^b_{nm}r^a_{mn}),
\end{align}
we obtain
\begin{align}
\Gamma_{bca;nm}
&=\frac{1}{2}\left(\d_cg_{ba;nm}+\d_ag_{bc;nm}-\d_bg_{ca;nm}\right)\notag\\
&=\frac{1}{2}{\rm Re}\left[\d_c(r^b_{nm}r^a_{mn})+\d_a(r^b_{nm}r^c_{mn})-\d_b(r^c_{nm}r^a_{mn})\right]\notag\\
&=\frac{1}{2}{\rm Re}\left[
r^b_{nm}r^a_{mn,c}+r^b_{nm,c}r^a_{mn}
+r^b_{nm}r^c_{mn,a}+r^b_{nm,a}r^c_{mn}
-r^c_{nm}r^a_{mn,b}-r^c_{nm,b}r^a_{mn}
\right]
\notag\\
&=\frac{1}{2}{\rm Re}\left[
2r^b_{nm}r^a_{mn,c}
+r^b_{nm}(r^c_{mn,a}-r^a_{mn,c})-(r^c_{nm,b}-r^b_{nm,c})r^a_{mn}
-r^c_{nm}(r^a_{mn,b}-r^b_{mn,a})
\right],
\end{align}
where we add $0=r^b_{nm}r^a_{mn,c}-r^b_{nm,c}r^a_{mn}-(r^b_{nm}r^a_{mn,c}-r^b_{nm,c}r^a_{mn})$ in the last line and use ${\rm Re}(r^b_{nm,a}r^c_{mn})={\rm Re}(r^c_{nm}r^b_{mn,a})$ for the last term.
Let us note that $r^c_{mn,a}-r^a_{mn,c}$ terms in the parenthesis are virtural transitions as can be seen from Eq.~\eqref{eq:rbmnc}.
Thus, we have
\begin{align}
\label{eq:multiband-connection}
\Gamma_{bca,nm}
&=
{\rm Re}\left[r^b_{nm}r^a_{mn,c}\right]
+\text{virtual transitions}.
\end{align}
Equivalently, we find that the matrix element of the circular shift current is given by
\begin{align}
{\rm Re}\left[r^b_{nm}r^a_{mn,c}-r^b_{nm,c}r^a_{mn}\right]
&=\Gamma_{bca;nm}-\Gamma_{acb;nm}
+\text{virtual transitions}.
\end{align}
An analogous proof for the complex part will be more challenging because it requires the calculation of the inverse quantum metric because $\tilde{\Gamma}_{bca}=\frac{1}{2}F_{bd}(g^{-1})^{de}\Gamma_{eca}$.

\end{widetext}

\section{Symmetry transformations of geometric quantities}

Under the transformation
\begin{align}
(t,x_a)\rightarrow (t',x_a')=((-1)^{s_T}t,{\cal M}_{ab}x_b),
\end{align}
the quantum geometric tensor and the quantum geometric connection transforms as
\begin{align}
Q_{ba}({\cal M}{\bf k})
&=
{\cal M}_{bb'}{\cal M}_{aa'}Q^*_{b'a'}({\cal M}^{-1}{\bf k}),\notag\\
C_{bca}({\bf k})
&=(-1)^{s_T}
{\cal M}_{bb'}{\cal M}_{cc'}{\cal M}_{aa'}C^*_{b'c'a'}({\cal M}^{-1}{\bf k}).
\end{align}
Equivalently, we can write it as
\begin{align}
g_{ba}({\bf k})
&=
{\cal M}_{bb'}{\cal M}_{aa'}g_{b'a'}({\cal M}^{-1}{\bf k}),\notag\\
F_{ba}({\bf k})
&=(-1)^{s_T}
{\cal M}_{bb'}{\cal M}_{aa'}F_{b'a'}({\cal M}^{-1}{\bf k}),\notag\\
\Gamma_{bca}({\bf k})
&=(-1)^{s_T}
{\cal M}_{bb'}{\cal M}_{cc'}{\cal M}_{aa'}\Gamma_{b'c'a'}({\cal M}^{-1}{\bf k}),\notag\\
\tilde{\Gamma}_{bca}({\bf k})
&=
{\cal M}_{bb'}{\cal M}_{cc'}{\cal M}_{aa'}\tilde{\Gamma}_{b'c'a'}({\cal M}^{-1}{\bf k}).
\end{align}

\section{Details of first-principles calculations}
\label{app:first-principles}
\begin{figure}[t] \centering
\includegraphics[width=7.5cm]{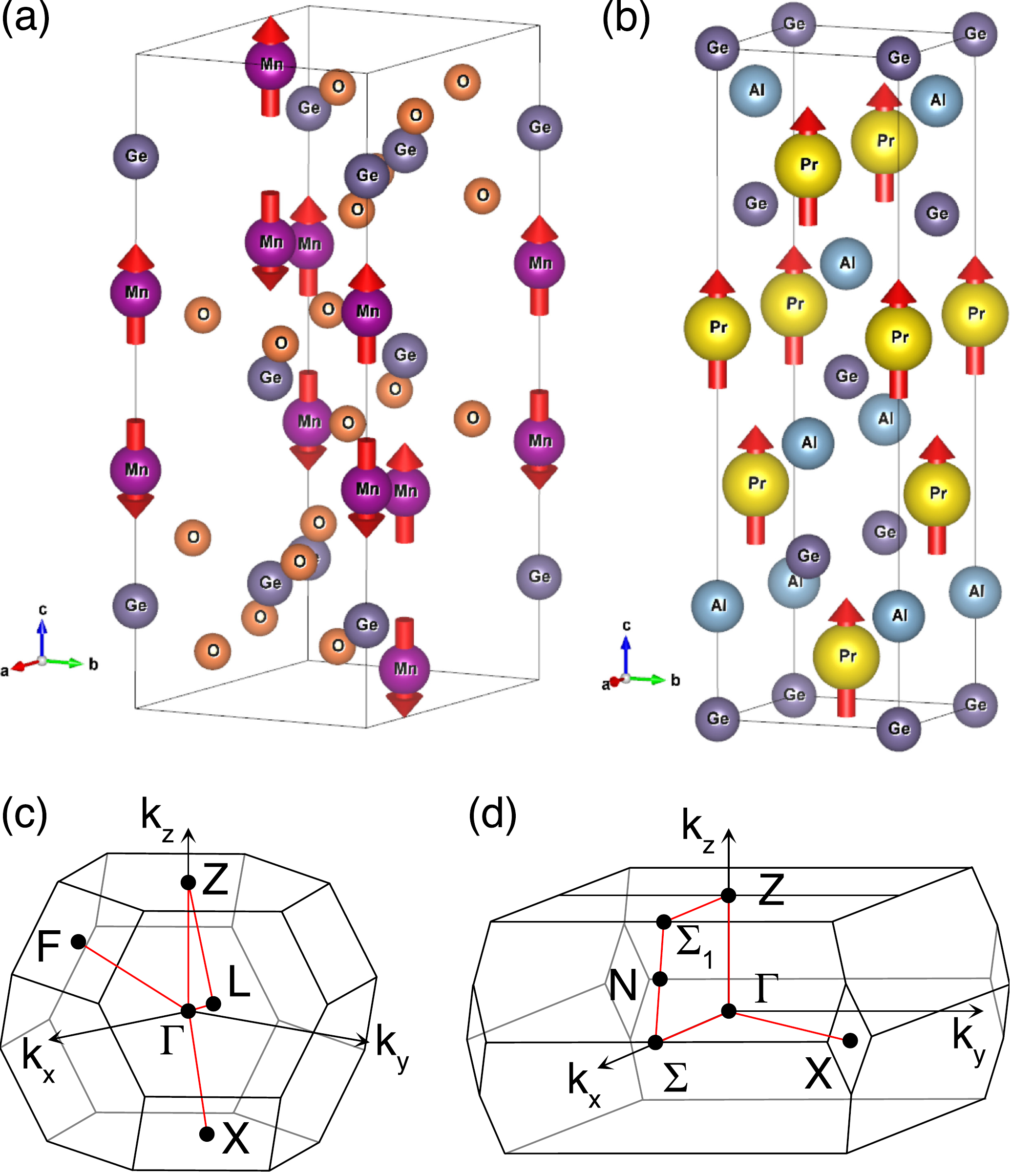}
\caption{Crystal and magnetic structures of (a) MnGeO$_3$ and (b) PrGeAl. The associated Brillouin zones
are shown in (c) and (d), respectively.
}
\label{fig:structure}
\end{figure}

Manganese germanate MnGeO$_3$ forms a rhombohedral ilmenite structure with a $P$ symmetric space group 
$R\overline{3}$~\cite{tsuzuki1974neutron} (see Fig.~\ref{fig:structure}a). The unit cell contains two formulae units (f.u.).
It becomes antiferromagnetic (AF) below 38 K with the same unit cell.~\cite{tsuzuki1974neutron}. 
Although the AF structure (Fig.~\ref{fig:structure}a)
breaks both $T$ and $P$ symmetries, AF MnGeO$_3$ with magnetic space group $-3'$ has the combined $PT$ symmetry.~\cite{xu2020high}.
PrGeAl crystallizes in the body-centered tetragonal structure (see Fig.~\ref{fig:structure}b)
with a broken $P$ symmetry space group $I4_1md$.~\cite{gladyshevskii2000crystal}
The unit cell also contains two f.u.
It becomes ferromagnetic (FM) at 16 K~\cite{sanchez2020observation}  and the FM structure (Fig.~\ref{fig:structure}b) has 
no $T$ symmetry nor $PT$ symmetry. 

\begin{figure}[t] \centering
\includegraphics[width=8.5cm]{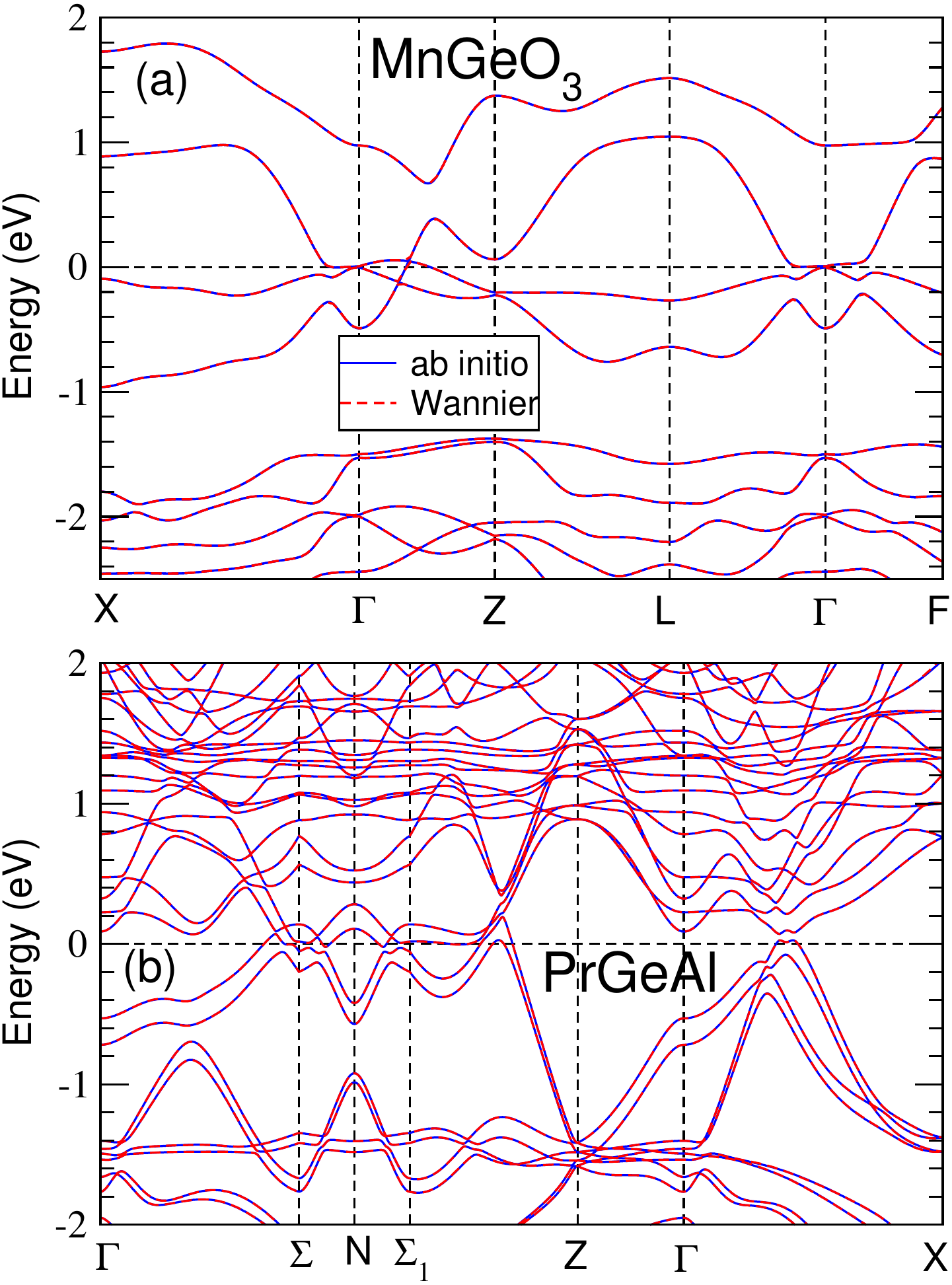}
\caption{Fully-relativistic band structures of (a) antiferromagnetic MnGeO$_3$ and (b) ferromagnetic PrGeAl.
The Fermi level ($E_F$) is at 0 eV.
}
\label{fig:BS}
\end{figure}


The electronic band structure and magnetic properties of MnGeO$_3$ and PrGeAl are calculated based on first-principles
density functional theory with the generalized gradient approximation (GGA)~\cite{perdew1996generalized}.
The experimental structural parameters for MnGeO$_3$~\cite{tsuzuki1974neutron} and PrGeAl~\cite{gladyshevskii2000crystal}
are used in the present calculations.
To better describe the Coulomb correlation among Mn 3$d$ electrons and also among Pr 4$f$ electrons, 
we adopt the GGA+$U$ scheme \cite{dudarev1998electron}.
Following the recent studies \cite{xu2020high,chang2018magnetic}, we use the effective $U$ value of 4.0 eV for
both Mn 3$d$ and Pr 4$f$ electrons.
The calculations are performed using the accurate projector-augmented wave (PAW) method~\cite{blochl1994projector},
as implemented in the Vienna {\it ab-initio} simulation package (VASP)~\cite{kresse1993ab,kresse1996efficient}.
The fully relativistic PAW potentials are adopted in order to include the spin-orbit coupling (SOC) effect.
Large plane-wave cutoff energies of 450 eV and 500 eV are used for MnGeO$_3$ and PrGeAl, respectively.
For the Brillouin zone integration, $k$-point meshes of 12 $\times$ 12 $\times$ 12 and 16 $\times$ 16 $\times$ 16
are used for MnGeO$_3$ and PrGeAl, respectively.
All the calculations are performed with an energy convergence within 10$^{-6}$ eV between the successive iterations.
\\
\\

Here we consider MnGeO$_3$ and PrGeAl, respectively, in the AF and FM states with the magnetic moments being parallel
to the $c$-axis. The calculated relativistic band structures of MnGeO$_3$ and PrGeAl are displayed in 
Fig.~\ref{fig:BS}(a) and Fig.~\ref{fig:BS}(b), respectively.  
They are both semimetals with low density of states at the Fermi level ($E_F$) 
of 0.315 states/eV/f.u. and  0.133 states/eV/f.u., respectively.
The calculated Mn spin magnetic moment in MnGeO$_3$ is 4.29 $\mu_B$
and that of Pr in PrGeAl is 1.89 $\mu_B$.
The calculated band structures agree well with that reported previously~\cite{xu2020high,chang2018magnetic}.

Nonlinear optical photocurrents are calculated based on the linear response formalism with the
independent particle approximation, as described above.
Specifically, the DC shift and injection photocurrent conductivity tensors are calculated
using Eq. (4) given in Sec. II.
Since a large number of $k$-points are needed to get accurate NLO responses~\cite{guo2004linear,wang2015nonlinear},
we use the efficient Wannier function interpolation method based on maximally localized Wannier functions
(MLWFs)~\cite{wang2006ab,marzari2012maximally,ibanez2018ab}.
For MnGeO$_3$, 68 MLWFs per unit cell of Mn $d$, Ge $p$  and O $p$ orbitals are constructed by fitting to the
GGA+U+SOC band structure. In PrGeAl, 76 MLWFs per unit cell of Pr $d$ and $f$ orbitals as well as Ge $p$  and Al $sp$ orbitals
are adopted. The band structures obtained by the Wannier
interpolation are nearly identical to that from the GGA+U+SOC calculations (see Fig.~\ref{fig:BS}).
The shift current conductivity tensors are then evaluated by
taking very dense $k$-point meshes of  200 $\times 200$ $\times$ 200 for MnGeO$_3$
and of 160 $\times 160$ $\times$ 220 for PrGeAl.
We find that the conductivity tensors obtained using such dense $k$-point meshes converge within a few percent.
Here we consider the "cold" semimetals, i.e., the Fermi-Dirac function in Eq. (4) is taken to be a step function.
Furthermore, the Dirac $\delta$ function is replaced by a Lorentzian function with broadenning width
of $\hbar\tau^{-1} = 10$ meV. Accordingly, in the injection conductivity calculations, we use
the relaxation time $\tau$ given by $\hbar\tau^{-1} = 10$ meV.


%

\end{document}